\documentclass[journal]{IEEEtran}
\usepackage{cite}
\usepackage{epsfig}
\usepackage[caption=false,font=footnotesize]{subfig}
\usepackage{graphicx}
\usepackage{amssymb}
\usepackage{array}
\newcolumntype{C}[1]{>{\centering\let\newline\\\arraybackslash\hspace{0pt}}m{#1}}
\usepackage{pifont}
\usepackage[cmex10]{amsmath, nccmath}
\usepackage{mathtools}
\usepackage{upgreek}
\usepackage{graphicx}
\usepackage{enumitem}
\usepackage{bm}
\newtheorem{theorem}{Theorem}

\newtheorem{lemma}{Lemma}
\newtheorem{definition}{Definition}

\newcommand{\norm}[1]{\left\lVert#1\right\rVert}

\newcommand{\E}[1]{\mathbb E\left[#1\right]}
\newcommand{\Prob}[1]{\mathbb P\left[#1\right]}

\newcommand{\mc}[1]{\mathcal{#1}}
\newcommand{\mb}[1]{\mathbf{#1}}
\newcommand{\mr}[1]{\mathrm{#1}}
\newcommand{\ms}[1]{\mathsf{#1}}

\newcommand{\thmref}[1]{Theorem~\ref{#1}}

\newcommand{\secref}[1]{Section~\ref{#1}}
\newcommand{\lemref}[1]{Lemma~\ref{#1}}

\newcommand{\appref}[1]{Appendix~\ref{#1}}

\newcommand{\figref}[1]{Fig.~\ref{#1}}

\DeclareMathOperator*{\argmax}{arg\,max}

\usepackage{hyperref}
\usepackage{url}
\usepackage{cases}
\hypersetup{colorlinks=true, linkcolor=blue, citecolor=blue, urlcolor = blue}
\usepackage{ifthen}
\newif\ifshowtodo
\showtodotrue

\renewcommand{\IEEEproofindentspace}{1\parindent}

\newcommand{\VersionLength}{long}
\providecommand{\ver}{\ifthenelse{\equal{\VersionLength}{long}}}
\allowdisplaybreaks
\begin{document}
\title{Variable-Length Feedback Codes \\ over Known and Unknown Channels \\ with Non-vanishing Error Probabilities }
\author{Recep Can Yavas and Vincent Y. F. Tan
\thanks{R. C. Yavas was with the Descartes Program, CNRS@CREATE, 138602, Singapore. He is now with the Department of Computer Science, National University of Singapore, 117417, Singapore (e-mail: ryavas@nus.edu.sg). V. Y. F. Tan is with the Department of Mathematics and Department of Electrical and Computer Engineering,
National University of Singapore, 119077, Singapore (e-mail: vtan@nus.edu.sg). This work is supported in part by the National Research Foundation, Prime Minister’s Office, Singapore under its CREATE programme. 

This paper was presented in part at ITW 2024 \cite{yavas2025VLF}.}}
\IEEEoverridecommandlockouts
\maketitle 
\begin{abstract}
We study variable-length feedback (VLF) codes with noiseless feedback for discrete memoryless channels. We present a novel non-asymptotic bound, which analyzes the average error probability and average decoding time of our modified Yamamoto--Itoh scheme. We then optimize the parameters of our code in the asymptotic regime where the average error probability $\epsilon$ remains a constant as the average decoding time $N$ approaches infinity. Our second-order achievability bound is an improvement of Polyanskiy \emph{et al.}'s (2011) achievability bound. We also develop a universal VLF code that does not rely on the knowledge of the underlying channel parameters. Our universal VLF code employs the empirical mutual information as its decoding metric and universalizes the code by Polyanskiy \emph{et al.} (2011).
We derive a second-order achievability bound for universal VLF codes. Our results for both VLF and universal VLF codes are extended to the additive white Gaussian noise channel with an average power constraint. The former yields an improvement over Truong and Tan's (2017) achievability bound. The proof of our results for universal VLF codes uses a refined version of the method of types and an asymptotic expansion from the nonlinear renewal theory literature.
\end{abstract}
\section{Introduction}
Feedback does not increase the capacity of memoryless channels \cite{shannon1956zero}. Yet, it simplifies the coding schemes that achieve the capacity \cite{horstein, schalkwijk}. For fixed-length codes, Wagner \emph{et al.} \cite{wagner2020} show that feedback improves the second-order achievable rate for discrete memoryless channels (DMCs) that have multiple capacity-achieving input distributions with distinct dispersions. 

The benefits of feedback are even more significant for variable-length feedback (VLF) codes, where the transmission stops at a random time depending on the noise realization. In his seminal work, Burnashev \cite{burnashev1976data} shows that the optimal error exponent (also known as the reliability function) for VLF codes over a DMC is given by
\begin{align}
    E(R) = \lim_{\epsilon \to 0} -\frac{1}{\E\tau } \log \epsilon  = C_1 \left(1 - \frac{R}{C} \right), \label{eq:burnashev}
\end{align}
where $C$ is the capacity of the DMC, $C_1 = \max_{x, x' \in \mc{X}} D(P_{Y|X = x} \| P_{Y|X = x'})$ is the Kullback--Leibler (KL) divergence between the conditional output distributions given the two most distinguishable input symbols, $R \in (0, C)$ is the rate, $\epsilon$ is the error probability, and $\E{\tau}$ is the average decoding time of the code. For any $R < C$, the error exponent in \eqref{eq:burnashev} is larger than that for fixed-length codes without feedback \cite{Gallager1965}. 
To achieve the optimal error exponent, Burnashev proposes a two-phase coding scheme, where in the \emph{communication phase}, the transmitter aims to increase the posterior of the transmitted message. If the largest posterior exceeds a threshold, the system goes into the \emph{confirmation phase}, where the decoder tries to verify the correctness of the estimate in the confirmation phase. 

Yamamoto and Itoh \cite{YamamotoItoh} propose an alternative scheme that achieves the optimal error exponent in \eqref{eq:burnashev}. 
Yamamoto and Itoh's scheme alternates between the communication and confirmation phases, each having fixed lengths, until a decision is made by the receiver. Any capacity-achieving fixed-length code can be used for the communication phase of the Yamamoto--Itoh scheme. In the confirmation phase, the transmitter transmits one of two control sequences, $(x_{\mr{A}}, \dots, x_{\mr{A}})$ and $(x_{\mr{R}}, \dots, x_{\mr{R}})$, where the first sequence indicates that the receiver should ``accept'' its current estimate, and the second sequence indicates that the receiver should ``reject'' its current estimate and start a new communication phase. The symbols $x_{\mr{A}}$ and $x_{\mr{R}}$ are chosen to be the two most distinguishable symbols in the sense that they achieve $C_1$. The receiver then constructs a (fixed-length) binary hypothesis test on the noisy versions of the control sequences and feeds its decision back to the transmitter. In \cite{chen2023}, Chen \emph{et al.} derive a non-asymptotic achievability bound for VLF codes with finite number of feedback instances; their code is a variant of the Yamamoto--Itoh scheme where the length of each communication and confirmation phase may be distinct. In \cite{berlin}, Berlin \emph{et al.} give an alternative proof to the converse of Burnashev's error exponent; their proof parallels the Yamamoto--Itoh scheme and reveals that communication and confirmation phases are implicit for any scheme that achieves the optimal error exponent.

Although   error exponent analysis elucidates how fast the error probability decays as the average decoding time $N \triangleq \E{\tau}$ grows to infinity, it does not explain the fundamental limit for a fixed error probability $\epsilon \in (0, 1)$ and a finite $N$ of our interest. To address this issue, Polyanskiy \emph{et al.} \cite{polyanskiy2011feedback} extend Burnashev's work to the regime with non-vanishing error probabilities and derive achievability and converse bounds on the logarithm of the maximum achievable codebook size $\log M^*(N, \epsilon)$ given an average decoding time $N$ and average error probability $\epsilon \in (0, 1)$. They show 
\begin{IEEEeqnarray}{rCl}
    \frac{N C}{1-\epsilon} - \log N + O(1)  \leq \log M^*(N, \epsilon) \leq \frac{N C}{1-\epsilon} + \frac{h_{\mr{b}}(\epsilon)}{1-\epsilon},\IEEEeqnarraynumspace \label{eq:pol}
\end{IEEEeqnarray}
where $h_{\mr{b}}(\epsilon) \triangleq - \epsilon \log \epsilon - (1-\epsilon) \log(1-\epsilon)$ is the binary entropy function. This result implies that the $\epsilon$-capacity is $\frac{C}{1-\epsilon}$, and the second-order term in the achievable rate is $O\left(\frac{\log N}{N}\right)$. To achieve the lower bound in \eqref{eq:pol}, they employ stop-feedback, which is a single bit of feedback that tells the transmitter whether to stop the transmission or to continue to transmit symbols. Polyanskiy \emph{et al.}'s scheme uses a stop-at-time-zero strategy, which decodes to an arbitrary message at time zero with probability $\epsilon_0 < \epsilon$, and with probability $1-\epsilon_0$, the scheme employs a code with an information-density threshold rule. Variants of Polyanskiy \emph{et al.}'s coding scheme with a finite number of feedback instances include \cite{kim2015VLF, williamson2015VLConvolution, vakilinia2016, yavas2021VLF, yavas2023VLF, Yang2022}. Some of the extensions of \cite{polyanskiy2011feedback} to multi-transmitter networks are \cite{yavas2023VLF, truong2018Journal}.
In \cite{naghshvar}, for symmetric binary-input channels, Naghshvar \emph{et al.} develop a deterministic, one-phase coding scheme that achieves the optimal error exponent in \eqref{eq:burnashev}. Their code has a novel encoder called the small-enough-difference (SED) encoder, which partitions the message set into two subsets at each time instance so that the probability difference between the two subsets is small enough. In \cite[Remark~4]{naghshvar2015EJS}, Naghshvar \emph{et al.} extend their work to arbitrary DMCs by introducing the maximum extrinsic Jensen-Shannon encoder and derive a non-asymptotic bound for their code. In \cite{yang2022SED}, Yang \emph{et al.} extend Naghshvar \emph{et al.}'s SED encoder to binary asymmetric channels (BACs) (i.e., channels with binary input and binary output), and derive refined non-asymptotic achievability bounds for the BAC and the binary symmetric channel (BSC).    

Since the exact channel statistics are not always available to the code designer, it is desirable to construct \emph{universal} codes in the sense that the DMC in use is known to belong to a certain family of DMCs (e.g., DMCs with known input and output alphabet sizes, BSCs with unknown flip probability), but the exact channel transition kernel $P_{Y|X}$ is unknown to both the transmitter and receiver. Naturally, we desire the performance of the universal code to be as close as possible to that of the non-universal code (e.g., the capacity, the error exponent). In \cite{goppa}, Goppa proposes to use the maximum (empirical) mutual information (MMI) decoder, which decodes to the message whose codeword has the maximum empirical mutual information with the received output sequence. Goppa shows that for DMCs, the MMI decoder attains capacity universally. In \cite[Th.~10.2]{csiszarbook}, Csisz\'ar and K\"orner  show that the random coding error exponent for constant-composition codes  is achieved universally by the MMI decoder. 
Universal channel coding is related to mismatched decoding, in which the decoder is fixed and potentially sub-optimal, and the goal is to optimize the codebook. This relationship stems from the fact that both mismatched decoding and universal coding attempt to address channel uncertainty (see \cite{scarlettmonograph} for a review of mismatched decoding). 
Merhav \cite{merhav2013} unifies the mismatched decoding and universal coding approaches, where he shows that for a given random coding distribution and a given class of metric decoders, their proposed generic universal decoder whose error probability is within a subexponential multiplicative factor of the best decoder in that class of decoders. Extensions of \cite[Th.~10.2]{csiszarbook} to the Gaussian channel with an unknown deterministic interference signal and to the Gaussian intersymbol interference channel appear in \cite{merhav93} and \cite{hule}, respectively.
In \cite{tchamkerten2006variable}, Tchamkerten and Telatar define universal VLF (UVLF) codes and show that Burnashev's error exponent is universally achieved over a family of BSCs with an unknown flip probability $p < \frac{1}{2}$ and over a family of Z channels with  unknown parameters. Their code is a universal Yamamoto--Itoh scheme tailored to the underlying BSC and Z channel families.
In \cite[Th.~3]{lomnitzfeder}, Lomnitz and Feder show that for DMCs, the rate that equals the empirical mutual information between input and output sequences is achievable universally in the VLF setting. In \cite[Th.~4]{lomnitzfeder}, they also show that for arbitrary continuous channels with an average power constraint, the rate $R = -\frac{1}{2} \log (1 - \hat{\rho}_{X^n Y^n}^2)$ is universally achievable in the VLF setting, where $\hat{\rho}_{X^n Y^n}^2$ is the empirical correlation between the input sequence $X^n$ and the output sequence $Y^n$. The quantity $-\frac{1}{2} \log (1 - \rho^2)$ corresponds to the mutual information of two Gaussian random variables with the correlation coefficient $\rho$. In~\cite{merhavfeder}, Merhav and Feder study the error exponents of universal decoding with an erasure option, where the trade-off between the probability of undetected error and the probability of erasure is considered; this problem is related to UVLF codes in the sense that at each time, the UVLF decoder chooses between decoding to the ``erasure'' option and decoding to a message. 

\subsection{Our Main Contributions}
In this paper, we study VLF and UVLF codes in the regime that the error probability $\epsilon \in (0, 1)$ is non-vanishing. For an arbitrary DMC with $C_1 < \infty$, equivalently, all entries of the channel transition kernel $P_{Y|X}$ are positive, we improve the second-order term in the lower bound in \eqref{eq:pol} for VLF codes from $-\log N $ to $-\frac{C}{C_1} \log N$. Our proposed VLF code is a modified Yamamoto--Itoh scheme with two communication and one confirmation phases, where each phase has a random stopping time, similar to the code in \cite{tchamkerten2006variable}. In \thmref{thm:nonasymp}, we derive a novel non-asymptotic achievability bound; in \thmref{thm:VLF}, we analyze the non-asymptotic bound to derive the asymptotic bound with the improved second-order term. 
In \thmref{thm:UVLF}, for UVLF codes, we derive an asymptotic achievability bound for an arbitrary DMC, where the second order term is $- \log N - \min\left\{\frac{|\mc{X}||\mc{Y}|}{2} , \left(|\mc{X}|-\frac{3}{2}\right)\left(|\mc{Y}|-\frac{3}{2}\right) + \frac{3}{4}\right\} \log N$. Our UVLF code universalizes Polyanskiy \emph{et al.}'s scheme in \cite{polyanskiy2011feedback} by replacing the information-density threshold rule in the communication phases with the empirical mutual information threshold rule. This empirical mutual information threshold rule is also used by Tchamkerten and Telatar \cite{tchamkerten2006variable}. Unlike in \cite{tchamkerten2006variable}, our UVLF code has a single phase. In the proof of \thmref{thm:UVLF}, we use the result in \cite[Th.~4.5]{woodroofebook} from the nonlinear renewal theory literature to bound the expected stopping time associated with the empirical mutual information. We also use the refined method of types from \cite{mardia2020} to get a tight tail probability bound for the empirical mutual information evaluated on a joint type formed from two independent sequences.

Theorems~\ref{thm:VLF_Gaussian} and \ref{thm:UVLF_Gaussian}, respectively, derive achievability bounds for VLF and UVLF codes to the Gaussian channel with an average power constraint. For UVLF codes over the Gaussian channel, we consider a scenario where the noise variance $\sigma_0^2$ of the channel is unknown to the transmitter and the receiver. Note that this model is equivalent to slow fading channel with a fixed an unknown fading factor and a known noise variance. For this problem, as our universal decoding metric, we employ the mutual information associated with the maximum likelihood estimator of the input-output pair $(X^n, Y^n)$ within the class of jointly Gaussian distributions, which equals $-\frac{1}{2} \log(1 - \hat{\rho}^2_{X^n Y^n})$, where $\hat{\rho}^2_{X^n Y^n}$ is the empirical correlation coefficient between $X^n$ and $Y^n$. This universal decoding metric is also used in \cite{lomnitzfeder}; a similar universal metric that also depends on the Gaussian input distribution is proposed in \cite[Example~2]{merhav2013}.  Our results here refine the achievability results of Lomnitz and Feder's first-order achievability bound in \cite[Th.~4]{lomnitzfeder}. Under the Gaussian input distribution $\mc{N}(0, P)$, the output of the Gaussian channel is $\mc{N}(0, P + \sigma_0^2)$, Unlike in the DMC case, this output distribution is a one-to-one function of the unknown channel parameter $\sigma_0^2$, thereby enabling the universalized Yamamoto--Itoh scheme for the Gaussian channel. Specifically, in the confirmation phase of the Yamamoto--Itoh scheme, we plug in the channel parameter estimated via the output sequence observed in the first communication phase without relying on whether the estimated message in the first communication phase is correct.

\subsection{Paper Organization}
The organization of the paper is as follows. \secref{sec:notation} defines the notation, \secref{sec:problemform} formulates the problems, \secref{sec:main} presents our main results, \secref{sec:Gaussian} extends our results to the Gaussian channel, and Sections~\ref{sec:proofVLF}--\ref{sec:proof_UVLF_Gaussian} contain the proofs. \secref{sec:conclusion} concludes the paper.

\section{Notation and Definitions} \label{sec:notation}
For $n \in \mathbb{N}$, we denote $[n] \triangleq \{1, \dots, n\}$ and the length-$n$ vector $x^n \triangleq (x_1, \dots, x_n)$. We denote the  collection of $M$ length-$n$ vectors as $\{x^n(1), \dots, x^n(M)\}$. The distribution of a random variable $X$ on an alphabet $\mc{X}$ is denoted by $P_X$. For a random variable $X$, we denote $X^+ \triangleq \max\{0, X\}$ and $X^- \triangleq -\min\{0, X\}$. The essential supremum of $X$ is defined as $\mr{ess} \sup (X) \triangleq \sup \{a \in \mathbb{R} \colon \Prob{X \geq a} > 0\}$. For any random variable $X$  with distribution $P_X$ and $\E{X} > 0$, we define the constant 
\begin{align}
    b(P_X) \triangleq \min\bigg\{ \frac{\E{(X^+)^2}}{\E{X}}, \mr{ess} \sup(X) \bigg\}.
\end{align}
The set of all distributions on $\mc{X}$ is denoted by $\mc{P}(\mc{X})$. The Gaussian random vector with mean $\bm{\mu}$ and covariance matrix $\ms{\Sigma}$ is denoted by $\mc{N}(\bm{\mu}, \ms{\Sigma})$. A random variable $X$ is called {\em arithmetic}\footnote{A {\em  lattice random variable with  span $h > 0$} is one in which  there exists some offset $a \in [0, h)$ such that $\Prob{X - a \in h \mathbb{Z}} = 1$ \cite{petrov1975}. An {\em arithmetic} random variable is a special case of a lattice random variable with zero offset. A random variable can be lattice but non-arithmetic.} with a span $h > 0$ if $\Prob{X \in h \mathbb{Z}} = 1$ and $h$ is the largest number that satisfies this condition; $X$ is called {\em non-arithmetic} if no such $h$ exists.

A DMC is defined by the single-letter channel transition kernel $P_{Y\vert X} \colon \mc{X} \to \mc{Y}$, where $\mc{X}$ and $\mc{Y}$ are the input and output alphabets. The DMC acts on each input symbol independently of others, i.e., $P_{Y^n|X^n}(y^n | x^n) = \prod_{i=1}^n P_{Y|X}(y_i | x_i)$ for all $x^n \in \mc{X}^n$ and $y^n \in \mc{Y}^n$. The set of all DMCs with input alphabet $\mc{X}$ and output alphabet $\mc{Y}$ is denoted by $\mc{P}(\mc{Y}|\mc{X})$. 

All logarithms have base $e$. The information density is defined as
\begin{align}
    \imath(x; y) \triangleq \log \frac{P_{Y|X}(y|x)}{P_Y(y)}, 
\end{align}
where the output distribution $P_Y$ is induced by a fixed input distribution $P_X$ and the DMC $P_{Y|X}$ (the dependence of the information density on $(P_X, P_{Y|X})$ is suppressed). The mutual information associated with  $P_X$ and  $P_{Y|X}$ is denoted as
\begin{align}
    I(P_X, P_{Y|X}) \triangleq \sum_{x \in \mathcal{X}, y \in \mathcal{Y}} P_{XY}(x, y) \log \frac{P_{Y|X}(y|x)}{ P_Y(y)}.
\end{align}
The capacity of a DMC $P_{Y|X}$ is 
\begin{align}
    C \triangleq \max \limits_{P_X \in \mc{P}(\mc{X})} I(P_X, P_{Y|X}).
\end{align}

The entropy of $P_X$ is denoted by $H(P_X)$, and the KL divergence between $P_X$ and $Q_X$ on the same alphabet $\mc{X}$ is denoted by $D(P_X \| Q_X)$. The error exponent \eqref{eq:burnashev} achieved when there are only 2 messages (which corresponds to the rate $R = 0$) is defined as
\begin{align}
    C_1 \triangleq \max_{x, x' \in \mc{X}} D(P_{Y|X = x} \| P_{Y|X = x'}). \label{eq:C1}
\end{align}

The empirical distribution (or \emph{type}) of a sequence $x^n \in \mathcal{X}^n$ is defined as
\begin{align}
    \hat{P}_{x^n}(x) \triangleq \frac{1}{n} \sum_{i = 1}^n 1\{x_i = x\}, \quad x \in \mc{X}. \label{eq:type}
\end{align}

The conditional type of a sequence $(x^n, y^n) \in \mc{X}^n \times \mc{Y}^n$ is defined as $\hat{P}_{y^n|x^n}(y|x) \triangleq \frac{\hat{P}_{x^n y^n}(x, y)}{\hat{P}_{x^n}(x)}$. The empirical mutual information associated with sequences $(x^n, y^n)$ is denoted by $I(\hat{P}_{x^n}, \hat{P}_{y^n|x^n})$. 
The set of length-$n$ types on an alphabet $\mc{X}$ is denoted by $\mc{P}_n(\mc{X}) \triangleq \{P_X \in \mc{P}(X) \colon n P_X(x) \in \mathbb{Z} \,\, \forall \, x \in \mc{X}\}.$ The type class of $P_X$ is defined as
$\mc{T}_n(P_X) \triangleq \{x^n \in \mc{X}^n \colon \hat{P}_{x^n} = P_X\}$.

We employ the standard $o(\cdot)$, $O(\cdot)$, $\Omega(\cdot)$, and $\Theta(\cdot)$ notations for asymptotic relationships of functions. 

\section{Problem Formulation} \label{sec:problemform}
We here formalize VLF and UVLF codes.  
\begin{definition}[VLF code {\cite[Def.~1] {polyanskiy2011feedback}}]
    \label{def:VLF}
    Fix $\epsilon \in (0, 1)$, $N > 0$, and a positive integer $M$. An $(N, M, \epsilon)$-VLF code comprises 
    \begin{enumerate}
        \item a common randomness random variable $U$ that has a finite alphabet $\mathcal{U}$ and an associated probability distribution $P_U$,\footnote{The need for common randomness arises because the code must simultaneously satisfy multiple constraints---in our case, an average error probability constraint and an average decoding time constraint.} (The realization $u$ of $U$ is revealed to the transmitter and receiver before the start of transmission to initialize the codebook.)
        \item encoding functions $\mathsf{f}_t \colon \mc{U} \times [M] \times \mc{Y}^{t-1} \times \mc{P}(\mc{Y}|\mc{X}) \to \mc{X}$ such that
        \begin{align}
            X_t = \ms{f}_t(U, W, Y^{t-1}, P_{Y|X}) \quad \forall t \in \mathbb{N},
        \end{align}
        where $W$ is the equiprobable message on $[M]$,

        \item a random stopping time $\tau \in \mathbb{N}$ of the filtration generated by $\{U, Y^{t}\}_{t = 0}^{\infty}$, which satisfies the average decoding time constraint
        \begin{align}
            \E{\tau} \leq N,
        \end{align}
        \item a decoding function $\ms{g}_{\mc{\tau}} \colon \mc{U} \times \mc{Y}^\tau \times \mc{P}(\mc{Y}|\mc{X}) \to [M]$ such that
        \begin{align}
            \hat{W} = \ms{g}_{\tau}(U, Y^\tau, P_{Y|X}), 
        \end{align}
        where $\hat{W}$ is the estimate of $W$. The estimate $\hat{W}$ must satisfy the average error probability constraint
        \begin{align}
        \Prob{\hat{W} \neq W} \leq \epsilon.
        \end{align}
    \end{enumerate}
\end{definition}

\begin{definition}[UVLF code]
    \label{def:UVLF}
    An $(N, M, \epsilon)$-UVLF code is defined similarly to an $(N, M, \epsilon)$-VLF code except that the encoding functions $\{\ms{f}_t\}_{t = 1}^{\infty}$ and the decoding function $\ms{g}_{\tau}$ can depend on the input and output alphabet sizes $|\mc{X}|$ and $|\mc{Y}|$ but not on the channel transition kernel $P_{Y|X}$.
\end{definition}

We define the maximum achievable codebook sizes $M^*(N, \epsilon)$ and $M_{\mr{U}}^*(N, \epsilon)$ as
\begin{align}
    M^*(N, \epsilon)&\triangleq \max\{M \in \mathbb{N} \colon\, \exists\, (N, M, \epsilon)\text{-VLF code}\} \label{eq:Mstar1} \\
    M_{\mathrm{U}}^*(N, \epsilon)
    &\triangleq \max\{M \in \mathbb{N} \colon \exists\, (N, M, \epsilon)\text{-UVLF code}\}. \label{eq:Mstar2}
\end{align}

\section{Main Result} \label{sec:main}
Our first result is a non-asymptotic achievability bound for VLF codes, where the channel transition kernel $P_{Y|X}$ is known.

\begin{theorem}\label{thm:nonasymp}
    Let $P_{Y|X}$ be the underlying DMC with $C_1 < \infty$ and $C > 0$. Fix a positive integer $M$, positive constants $\gamma_1 < \gamma_2$, $a_{\mr{A}}$, and $a_{\mr{R}}$, $\epsilon_0 \in (0, 1)$, and a capacity-achieving input distribution $P_X$. 
    Define
    \begin{align}
        (x_{\mr{A}}, x_{\mr{R}}) \triangleq \argmax_{(x, x') \in \mc{X}^2} D(P_{Y|X = x} \| P_{Y|X = x'}). \label{eq:xAxR}
    \end{align}
    There exists an $(N, M, \epsilon)$-VLF code with 
    \begin{align}
        N &\leq (1-\epsilon_0) N' \label{eq:NboundNpr}\\
        \epsilon &\leq \epsilon_0 + (1- \epsilon_0) \epsilon', \label{eq:epsbound}
    \end{align}
    where
    \begin{align}
        \epsilon' &= (M-1) \left(\exp\{-(\gamma_1 + a_{\mr{A}})\} + \exp\{-\gamma_2\} \right) \label{eq:epsprime_bound} \\
        N' &= \frac{\gamma_1 + b}{C} \notag \\
        &+ ((M-1) \exp\{-\gamma_1\} + \exp\{-a_{\mr{R}}\}) \frac{\gamma_2 - \gamma_1 + b}{C} \notag \\
        &+\frac{a_{\mr{A}} + b_{\mr{A}}}{D(P_{Y|X = x_{\mr{A}}} \| P_{Y|X = x_{\mr{R}}})} \notag \\
        &+ (M-1) \exp\{-\gamma_1\} \frac{a_{\mr{R}} + b_{\mr{R}}}{D(P_{Y|X = x_{\mr{R}}} \| P_{Y|X = x_{\mr{A}}})} \label{eq:Nprime_bound}\\
        b &= b(P_Z), \quad b_{\mr{A}} = b(P_{Z_{\mr{A}}}), \quad
        b_{\mr{R}} = b(P_{Z_{\mr{R}}}),
    \end{align}
    $(X, Y) \sim P_X P_{Y|X}$,  $Y_{\mr{A}} \sim P_{Y|X = x_{\mr{A}}}$, $Y_{\mr{R}} \sim P_{Y|X = x_{\mr{R}}}$, $Z = \imath(X; Y)$, $Z_{\mr{A}} = \log \frac{P_{Y|X = x_{\mr{A}}}(Y_{\mr{A}})}{P_{Y|X = x_{\mr{R}}}(Y_{\mr{A}})}$, and $Z_{\mr{R}} = \log \frac{P_{Y|X = x_{\mr{R}}}(Y_{\mr{R}})}{P_{Y|X = x_{\mr{A}}}(Y_{\mr{R}})}$.
\end{theorem}
    
\begin{IEEEproof}
    See \secref{sec:proof_nonasymp}.
\end{IEEEproof}

The proposed coding scheme to prove \thmref{thm:nonasymp} is a variant of the Yamamoto--Itoh scheme \cite{YamamotoItoh} and is modified from Tchamkerten's and Telatar's VLF coding scheme \cite{tchamkerten2006variable}, which is designed for unknown channels. Our code that achieves \eqref{eq:NboundNpr}--\eqref{eq:epsbound} is similar to the code in \cite{chen2023} in limiting the number of phases to a finite integer, but differs from it as each phase in our code has a random stopping time. Our code has two communication phases (C1 and C2) and one confirmation phase (HT), where the HT phase is between the C1 and C2 phases. 
We combine the Yamamoto--Itoh scheme with the stop-at-time-zero strategy used in \cite{polyanskiy2011feedback}, in which the code stops and decodes to an arbitrary message at time zero with probability $\epsilon_0$ and employs the Yamamoto--Itoh scheme with probability $1-\epsilon_0$. Decoding occurs either at time zero, or at the end of the HT phase, or at the end of the C2 phase. At large average decoding times $N$, the stop-at-time-zero strategy with a non-zero $\epsilon_0$ improves the achievable rate, and asymptotically achieves the $\epsilon$-capacity $\frac{C}{1-\epsilon}$. This strategy is also employed in \cite{yavas2021VLF, yavas2023VLF, truong2017gaussian}. The details of our code design appears in \secref{sec:codingscheme}.

Naghshvar \emph{et al.} \cite[Remark 7]{naghshvar2015EJS}  prove that there exists an SED encoder with parameters $(N', M, \epsilon')$ such that\footnote{The bound in \eqref{eq:nag} is achieved by the MaxEJS encoder for general DMCs \cite[Remark~4]{naghshvar2015EJS} and by the SED encoder for symmetric binary-input DMCs \cite[Remark~7]{naghshvar2015EJS}. The SED encoder is the computationally-efficient version of the MaxEJS encoder, tailored to a specific class of binary-input channels.}
\begin{align}
    N' \leq \frac{\log M + \log \log \frac{M}{\epsilon'}}{C} + \frac{\log \frac{1}{\epsilon'}}{C_1} + G, \label{eq:nag}
\end{align}
where 
\begin{align}
    G &\triangleq \frac{6 \cdot (4C_2)^2 + C}{C C_1} \\
    C_2 &\triangleq \max_{y \in \mc{Y}} \frac{\max_{x \in \mc{X}} P_{Y|X}(y|x)}{\min_{x \in \mc{X}} P_{Y|X}(y|x)}.
\end{align}
Here, the constant $G$ depends only on the DMC $P_{Y|X}$.
Although the bound in \eqref{eq:nag} is sufficient to show that the SED encoder achieves Burnashev's optimal error exponent in \eqref{eq:burnashev}, its non-asymptotic performance for moderate values of $N'$ such as $N' \in [10^2, 10^3]$ is poor. For example, for the binary-input ternary-output DMC that is the cascade of a BSC with flip probability 0.11 and a binary erasure channel (BEC) with erasure probability 0.2, $G$ roughly equals $1.73 \times 10^4$. To compare the performance of the SED encoder with that of \thmref{thm:nonasymp}, we consider the combination of the SED encoder with the stop-at-time-zero strategy. Then, there exists an $(N, M, \epsilon)$-VLF code such that
\begin{align}
    N &\leq (1-\epsilon_0) N' \label{eq:Nnag} \\
    \epsilon &\leq \epsilon_0 + (1-\epsilon_0) \epsilon', \label{eq:epsnag}
\end{align}
where $(N', \epsilon')$ satisfies \eqref{eq:nag}.

\begin{figure}[htbp] 
    \centering
    \subfloat[]{
    	\includegraphics[width=0.48\textwidth]{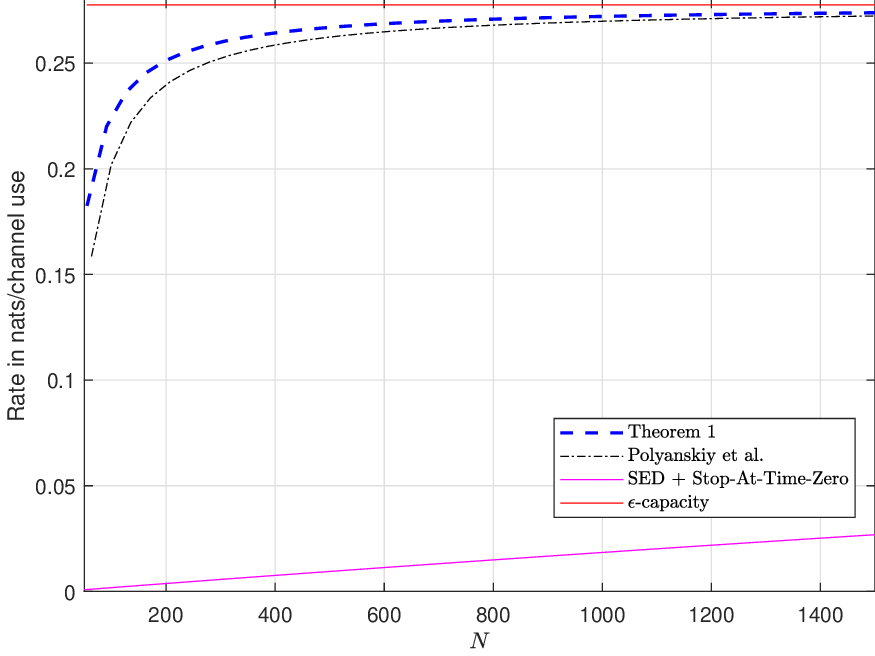}
    	\label{fig:BITO}} \\
    
    \subfloat[]{	
    	\includegraphics[width=0.48\textwidth]{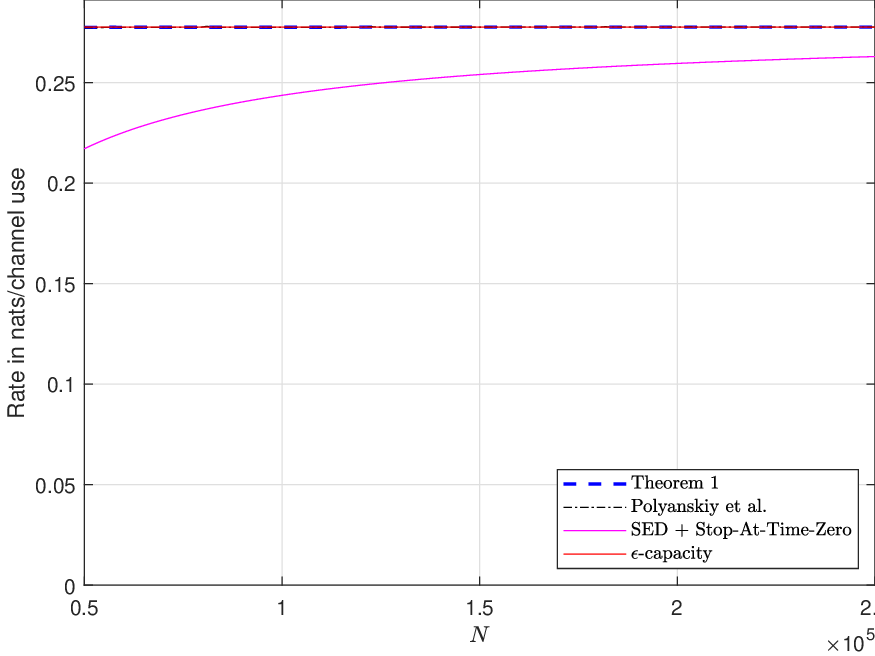}
}
     
    \caption{\label{fig:figure} Achievable rates over the cascade of a BSC with flip probability 0.11 and a BEC with erasure probability 0.2 are shown. The target error probability is $\epsilon = 10^{-3}$ for both figures. The average decoding time $N$ ranges in [100, 1500] for (a) and in $[5 \times 10^4, 2.5 \times 10^5]$ for (b).}
    
\end{figure}

\begin{figure}[htbp] 
    \centering
    
    	\includegraphics[width=0.48\textwidth]{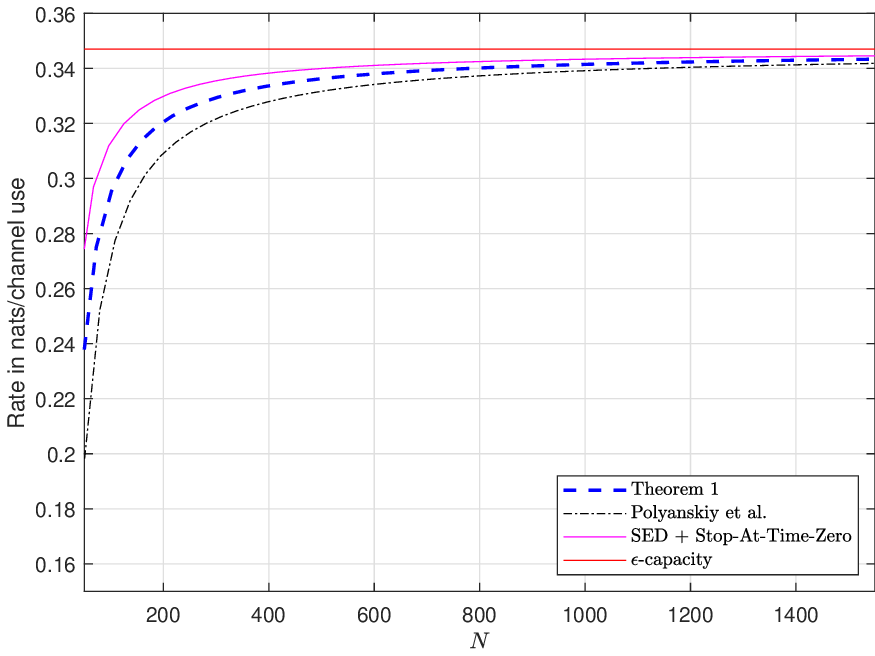}
     
    \caption{\label{fig:BSC} Achievable rates over the BSC with flip probability 0.11 are shown. The target error probability is $\epsilon = 10^{-3}$, and the average decoding time is $N \in [100, 1500]$.}
    
\end{figure}

In Figs.~\ref{fig:figure} and~\ref{fig:BSC},   achievable rates are presented for our VLF code (\thmref{thm:nonasymp}) where the parameters of the code are optimized numerically, the
SED encoder \cite{naghshvar2015EJS} combined with the stop-at-time-zero strategy given in \cite{polyanskiy2011feedback}, and Polyanskiy \emph{et al.}'s VLSF codes in \cite{polyanskiy2011feedback}. The two channels selected are the cascade of BSC(0.11) and BEC(0.2) for \figref{fig:figure} and BSC(0.11) for \figref{fig:BSC}.
Because the value of $G$ is too large relative to the values of $N$ displayed, after optimizing the parameters $(N', \epsilon', \epsilon_0)$, the $G$ term dominates the right-hand side of \eqref{eq:nag}. On other hand, the first two terms on the right-hand side of \eqref{eq:nag} dominate in the asymptotic case where $M$ approaches infinity, and $\epsilon$ approaches zero. This is reflected in \figref{fig:figure}, where the performance of the SED encoder is much worse than \thmref{thm:nonasymp}.

Yang \emph{et al.} \cite[Th.~7]{yang2022SED} derive a refined bound specific to BSCs by analyzing the performance of the SED encoder; in particular, their result improves the bound on $G$ from $\approx 1.11 \times 10^4$ to $\approx 5.41$ for BSC(0.11). Similarly, in \figref{fig:BSC}, we compare \thmref{thm:nonasymp} with the refined bound in \cite[Th.~7]{yang2022SED} combined with \eqref{eq:Nnag}--\eqref{eq:epsnag} for BSC(0.11).
We observe that Yang \emph{et al.}'s bound is slightly tighter than ours for the given values of $N$. We here note that our Yamamoto--Itoh-type code uses much less feedback compared to  that of the SED encoder. Specifically, our code employs stop-feedback at all time instants except at the end of C1 and HT phases (see Table~\ref{tab}, below); the SED encoder uses the whole output sequence $Y^n$ to determine the value of the transmitted symbol at time $n + 1$.\footnote{If feedback were noisy, the power allocated for the ``continue'' signal of the stop-feedback code would be much less than the one for the ``stop'' signal because the ``stop'' signal needs to be transmitted only once. For the SED encoder, the power allocation of the feedback signal would be uniform over time.}
It remains an open question whether the non-asymptotic bound in~\eqref{eq:nag} can be improved further for general DMCs to make it competitive.

Our second result is a second-order achievability bound for VLF codes, where the error probability $\epsilon \in (0, 1)$ is fixed as the average decoding time $N$ approaches infinity.
\begin{theorem} \label{thm:VLF}
Let $P_{Y|X}$ be the underlying DMC with $C > 0$ and $C_1 < \infty$. Then, 
    \begin{IEEEeqnarray}{rCl}
        \log M^*(N, \epsilon) \geq \frac{NC}{1-\epsilon} - \frac{C}{C_1} \log N -\log \log N + O(1). \IEEEeqnarraynumspace \label{eq:VLFeq}
    \end{IEEEeqnarray}
\end{theorem}
\begin{IEEEproof}
    The proof of \thmref{thm:VLF} follows from carefully choosing the parameters $\gamma_1, \gamma_2, a_{\mr{A}}$, $a_{\mr{R}}$, and $\epsilon_0$ in \thmref{thm:nonasymp}, and appears in \secref{sec:proofVLF_2}.
\end{IEEEproof}

Since $C \leq C_1$, \thmref{thm:VLF} improves the second-order term in \cite[eq.~(18)]{polyanskiy2011feedback} given in the lower bound in \eqref{eq:pol} from $-\log N$ to $- \frac{C}{C_1} \log N$. The achievability bound in \cite[eq.~(18)]{polyanskiy2011feedback} employs stop-feedback while our Yamamoto--Itoh scheme employs stop-feedback and also sends a $\lceil \log_2 M \rceil$-bits of feedback at the end of the first communication phase. 
The improvement in the second-order term results from the fact  that the error probability of our scheme is dominated by the error probability terms due to the confirmation phase and the second communication phase, whose average length scales as the logarithm of the average length of the first communication phase.
For general DMCs, the non-asymptotic bound in \cite[Remark~4]{naghshvar2015EJS} for Naghshvar \emph{et al.}'s MaxEJS encoder achieves a second-order term $-\left(\frac{C}{C_1} + 1\right) \log N$ when combined with the stop-at-time-zero strategy. To the best of our knowledge, \thmref{thm:VLF} yields the best asymptotic achievability bound for VLF codes with non-vanishing error probabilities over general DMCs. For BSCs and BACs, Yang \emph{et al.}'s bounds from \cite[Th.~4~and~7]{yang2022SED} recover \eqref{eq:VLFeq} with $-\log \log N + O(1)$ improved to $O(1)$ when combined with the stop-at-time-zero strategy.  It remains open to close the gap between the achievability bound in \thmref{thm:VLF} and the converse bound on the right-hand side of \eqref{eq:pol}.

The third result is a second-order achievability bound for universal VLF codes, where the DMC $P_{Y|X}$ is unknown but a capacity-achievability input distribution $P_X$ is known. We assume that the error probability $\epsilon \in (0, 1)$ is non-vanishing as the average decoding time $N$ approaches infinity.
\begin{theorem} \label{thm:UVLF}
Assume that a capacity-achieving distribution of the DMC $P_{Y|X}$ is known. Assume that $C > 0$ and $C_1 < \infty$.  Then, 
    \begin{align}
        &\log M^*_{\mr{U}}(N, \epsilon) \geq \frac{NC}{1-\epsilon} - \log N \notag \\
        &- \min\left\{\frac{|\mc{X}||\mc{Y}|}{2} , \left(|\mc{X}|-\frac{3}{2}\right)\left(|\mc{Y}|-\frac{3}{2}\right) + \frac{3}{4}\right\} \log N \notag \\
        & + o(\log \log N). \label{eq:UVLF}
    \end{align}
    In the case where $P_{Y|X}$ is known to be a BSC with an unknown flip probability $p \in (0, 1) \setminus \{\frac{1}{2}\}$, \eqref{eq:UVLF} is improved to
    \begin{align}
        \log M_{\mr{U}}^*(N, \epsilon) &\geq \frac{NC}{1-\epsilon} - \frac{3}{2} \log N   + o(\log \log N). \label{eq:BSCUVLF}
    \end{align}
    For an arbitrary (not necessarily capacity-achieving) random coding input distribution $P_X$, our universal code achieves the right-hand side of \eqref{eq:UVLF} with the capacity $C$ replaced by the mutual information $I(P_X, P_{Y|X})$.
    
\end{theorem}

\begin{IEEEproof}
    The proof of \thmref{thm:UVLF} differs from the proof of \thmref{thm:VLF} in two main ways. First, we bound $\Prob{\tau_2 \leq n_2} \leq \sum_{n = 1}^{n_2} \Prob{n I(\hat{P}_{\bar{X}^n}, \hat{P}_{Y^n|\bar{X}^n}) > \gamma}$ using the refined method of types bound from \cite[Th.~3]{mardia2020} and a refined bound on $\E{\exp\{n I(\hat{P}_{\bar{X}^n}, \hat{P}_{Y^n|\bar{X}^n})\}}$ combined with Markov's inequality. Here, $n_2$ is a suitably chosen constant and $(\bar{X}^n, Y^n) \sim P_X^n P_Y^n$. The third term on the right-hand side of \eqref{eq:UVLF} results from the additional multiplicative factor of $n_2^{d}$ in the bound on $\Prob{\tau_2 \leq n_2}$ compared to \eqref{eq:taumibound}, where $-d$ is the coefficient of the third term on the right-hand side of \eqref{eq:UVLF}. Second, to bound the expected stopping time, we use \cite[Th.~4.5]{woodroofebook} from the nonlinear renewal theory, which bounds the expected value of the stopping time $\tau = \inf\{n \geq 1\colon n g\left(\frac{1}{n} S_n\right) > \gamma\}$, where $S_n$ is a sum of $n$ i.i.d. vectors, and $g$ is a sufficiently smooth function. After we apply this result with $g$ being the mutual information function and $S_n$ being the empirical joint distribution of $\mb{c}^n(1)$ and $Y^n$, we get
\begin{align}
    \E{\tau} \leq \E{\tau_1} \leq \frac{\gamma}{C} + O(1).
\end{align}
This implies that the expected stopping time associated with the empirical mutual information admits the same asymptotic bound associated with the information density up to an $O(1)$ gap (see \lemref{lem:tau}, below). The analysis in \cite{tchamkerten2006variable} yields the bound $\E{\tau} \leq \frac{\gamma}{C} (1 + o(1))$ as $\gamma \to \infty$, which is not sharp enough to prove the $\log N$ scaling of the second-order term in \eqref{eq:UVLF}. To prove \eqref{eq:BSCUVLF}, we replace the decoding metric by $n(\log 2 - H(\hat{P}_{Z^n(m)}))$, where $Z_i(m) = 1\{Y_i \neq \mb{c}_i(m)\}$ is the Hamming distance between $Y_i$ and $\mb{c}_i(m)$, for $i \in [n]$. The proof of \thmref{thm:UVLF} is given in \secref{sec:proofUVLF}.
\end{IEEEproof}

\begin{figure}[htbp] 
   \centering
    \subfloat[]{
    	\includegraphics[width=0.48\textwidth]{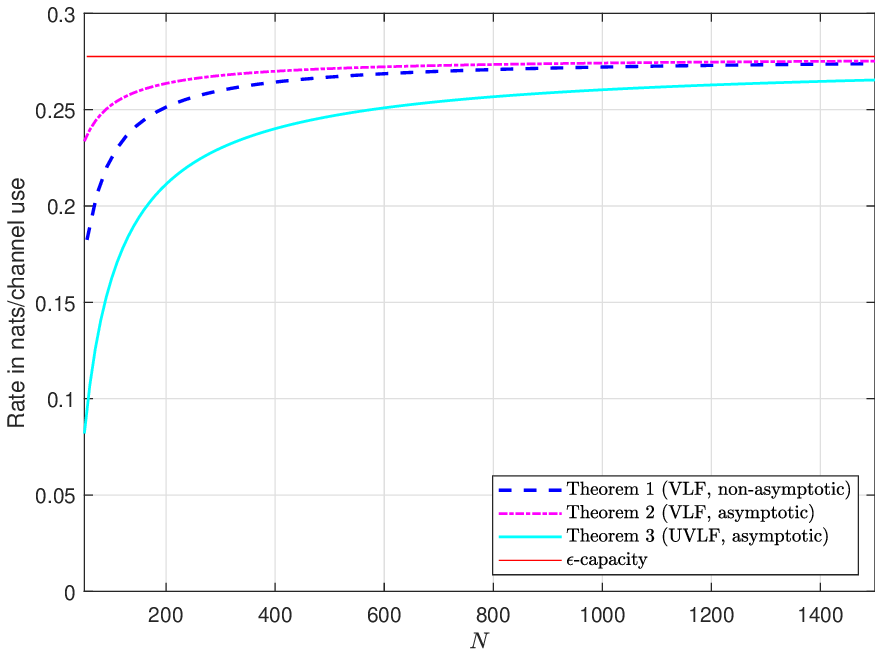}
    	\label{fig:Th23BITO}} \\
    
    \subfloat[]{	
    	\includegraphics[width=0.48\textwidth]{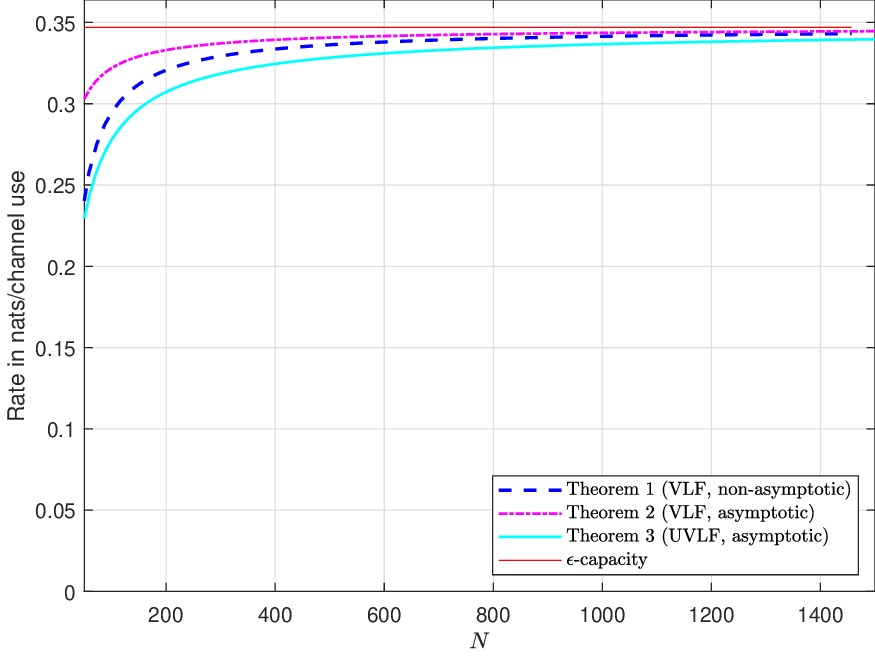}
}
     
    \caption{\label{fig:Th23} Asymptotic expansions in Theorems~\ref{thm:VLF}--\ref{thm:UVLF} (a) over the cascade of BSC(0.11) and BEC(0.2) and (b) over BSC(0.11) are shown. The average decoding time is $N \in [100, 1500]$, and the error probability is $\epsilon = 10^{-3}$.}
    
\end{figure} 

In \figref{fig:Th23},  achievable rates in \thmref{thm:VLF} for VLF codes and in \thmref{thm:UVLF} for UVLF codes are presented over (a) the cascade of BSC(0.11) and BEC(0.2) and (b) over BSC(0.2). For \thmref{thm:VLF}, the $O(1)$ term in \eqref{eq:VLFeq} is ignored, and for \thmref{thm:UVLF}, the $o(\log \log N)$ term in \eqref{eq:UVLF} is ignored. For VLF codes,  as we expect, the gap between the curves associated with the non-asymptotic bound (\thmref{thm:nonasymp}) and the asymptotic bound (\thmref{thm:VLF}) diminishes. We lack a non-asymptotic counterpart of \thmref{thm:UVLF} for UVLF codes because a non-asymptotic version of \lemref{lem:nonlinear}, which we use to bound the expected stopping time for UVLF codes, is not available in the literature, and appears challenging to derive.

The coding scheme to achieve the right-hand side of \eqref{eq:UVLF} universalizes Polyanskiy \emph{et al.}'s  single-phase coding scheme in \cite{polyanskiy2011feedback}, which combines the information density threshold rule by the stop-at-time-zero strategy. Since the probability transition kernel $P_{Y|X}$ is unknown to the code designer, we replace the information density $\imath(\mb{c}^{n}(m); Y^n)$ by the empirical mutual information $n I(\hat{P}_{\mb{c}^n(m)}, \hat{P}_{Y^n | \mb{c}^n(m)})$ and choose the threshold $\gamma$ and the probability to stop at time zero, $\epsilon_0$, as a function of $M, |\mc{X}|, |\mc{Y}|$, and $\epsilon$ only. The joint empirical distribution $\hat{P}_{\mb{c}^n(m)} \times \hat{P}_{Y^n|\mb{c}^n(m)}$ coincides with the maximum likelihood estimator within the family of distributions with alphabet $\mc{X} \times \mc{Y}$. 

In \cite{tchamkerten2006variable}, Tchamkerten and Telatar propose a variation of Yamamoto--Itoh scheme tailored to BSCs  with  crossover probability $p \in (0, \frac{1}{2})$. Specifically, in the HT phase, they use a statistics that is independent of $p$, i.e., the difference between the number of 1's and the number of 0's observed in the HT phase. Although the resulting reliability function associated with their universal sequential HT phase is optimal, both the average stopping time and the error probability exponent depend on the unknown flip probability of the BSC. In result, both the rate that the code operates at and the error probability of the code in \cite{tchamkerten2006variable} depend on the unknown capacity of the BSC.\footnote{In eq. (110) and (113) of \cite{tchamkerten2006variable}, it is shown that the error probability associated with the HT phase is bounded by $\exp\{-\gamma \frac{D(\mr{Bern}(p) \| \mr{Bern}(1-p))}{1-2p} \}$, and the average stopping time for the HT phase is $\frac{\gamma}{1 - 2p}(1 + o(1))$, where $\gamma$ is the threshold of the test chosen by the code designer. Although this test achieves the universally optimal error exponent $D(\mr{Bern}(p) \| \mr{Bern}(1-p))$ for $p \in (0, \frac{1}{2})$, it is not useful for our setup since we seek to bound the error probability of the test as a function of the threshold $\gamma$ only.} Since our goal is to control the error probability of the universal code regardless of the channel in use, we instead appeal to the single-phase scheme, which is essentially obtained by removing the HT phase from the Yamamoto--Itoh scheme.

\section{Extension to the Gaussian Channel} \label{sec:Gaussian}
The output of a memoryless Gaussian channel of blocklength $n$ in response to the input $X^n \in \mathbb{R}^n$ is 
        \begin{align}
            Y^n = X^n + Z^n, \label{eq:pointchannel}
        \end{align}
        where $Z_1, \ldots, Z_n$ are drawn i.i.d. from $\mathcal{N}(0, \sigma_0^2)$, independent of $X^n$, and $\sigma_0^2 > 0$ is the noise variance.
        Define the signal-to-noise ratio 
        \begin{align}
        S \triangleq \frac{P}{\sigma_0^2}, \label{eq:SNR}
        \end{align}
        where $P$ is the per-symbol average power constraint (see, \eqref{eq:powercons}).
        The capacity-cost function of the Gaussian channel is defined as
        \begin{align}
            C(S) &\triangleq \frac{1}{2} \log  (1 + S).
        \end{align}
        The analog of the quantity $C_1$ in \eqref{eq:C1} for the Gaussian channel is defined as
        \begin{align}
            C_1(S) &\triangleq  D(P_{Y|X = \sqrt{P}} \| P_{Y|X = -\sqrt{P}}) \\
            &= D(\mc{N}(\sqrt{P}, \sigma_0^2) \| \mc{N}(-\sqrt{P}, \sigma_0^2)) \\
            &= 2 S.
        \end{align}

\begin{definition} \label{def:VLF_Gaussian}
    An $(N, M, \epsilon, P)$-VLF code and an $(N, M, \epsilon, P)$-UVLF code are defined similarly to Definitions~\ref{def:VLF} and \ref{def:UVLF} with the addition of average power constraints
    \begin{align}
        \E{\sum_{t = 1}^{\tau} \ms{f}_t(U, W, Y^{t-1}, P_{Y|X})^2} \leq N P \label{eq:powercons} \\
        \E{\sum_{t = 1}^{\tau} \ms{f}_t(U, W, Y^{t-1})^2} \leq N P,
    \end{align}
    respectively, where $\tau$ is the random decoding time, and $P$ is the average power per symbol. We define the maximum achievable codebook sizes $M^*(N, \epsilon, P)$ and $M^*_{\mr{U}}(N, \epsilon, P)$ similarly to \eqref{eq:Mstar1}--\eqref{eq:Mstar2}.
\end{definition} 

The average power constraint in \eqref{eq:powercons} is introduced in \cite{truong2018Journal} for variable-length stop-feedback codes for the Gaussian multiple-access channel. 

The following achievability bound extends Theorem~\ref{thm:VLF} to the Gaussian channel with an average power constraint. 

\begin{theorem} \label{thm:VLF_Gaussian}
    Let $\sigma_0^2 > 0$ be the noise variance of the Gaussian channel, and let $P$ be the average power constraint. Recall that the signal-to-noise ratio is $S = \frac{P}{\sigma_0^2}$.
    For the Gaussian channel with the noise variance $\sigma_0^2$,
    \begin{align}
        \log M^*(N, \epsilon, P) &\geq \frac{N C(S)}{1-\epsilon} - \frac{C(S)}{C_1(S)} \log N \notag \\
        &\quad - \log \log N + O(1).
    \end{align}
\end{theorem}
\begin{IEEEproof}
    \thmref{thm:VLF_Gaussian} is proved using our Yamamoto--Itoh scheme described in \eqref{eq:codedist}--\eqref{eq:taumibound}. During the communication phases, i.e., the C1 and C2 phases, the input symbols are drawn i.i.d. from the Gaussian distribution $\mc{N}(0, P)$, which satisfies the average power constraint in \eqref{eq:powercons}. During the HT phase, the transmitter sends either $(\sqrt{P}, \sqrt{P}, \dots)$ or $(-\sqrt{P}, -\sqrt{P}, \dots)$ to accept or reject the receiver's initial estimate, respectively. Since the techniques used in the proof of \thmref{thm:nonasymp} applies to continuous random variables, \thmref{thm:nonasymp} applies to the Gaussian channel with $P_X = \mathcal{N}(0, P)$, $P_{Y|X = x} = \mc{N}(x, \sigma_0^2)$, and $C = C(S)$. \thmref{thm:VLF_Gaussian} follows by following the same steps as in the proof of \thmref{thm:VLF}.
\end{IEEEproof}

Since $C_1(S) < C(S)$ for every $S > 0$, \thmref{thm:VLF_Gaussian} improves the second-order term in the achievability bound in \cite[Th.~1]{truong2016gaussian} from $-\log N$ to $-\frac{C(S)}{C_1(S)} \log N$. As an analog to the DMC scenario in \eqref{eq:pol}, in \cite[Th.~1]{truong2016gaussian}, Truong and Tan show the converse bound\footnote{Truong and Tan prove the bound only for stop-feedback codes, which are a subset of VLF codes; however, the same proof applies to VLF codes as well.}
\begin{align}
    \log M^*(N, \epsilon, P) &\leq \frac{N C(S)}{1-\epsilon} + \frac{h_{\mr{b}}(\epsilon)}{1-\epsilon}. \label{eq:Gaussianconv}
\end{align}
Similar to the DMC case, there is a gap of $O(\log N)$ between the maximum achievable codebook sizes in the best achievability (\thmref{thm:VLF_Gaussian}) and converse (eq. \eqref{eq:Gaussianconv}) bounds; closing this gap remains an open problem. 

The following achievability bound for UVLF codes extends Theorem~\ref{thm:UVLF} to the Gaussian channel with an average power constraint. 

\begin{theorem} \label{thm:UVLF_Gaussian}
   Under the settings of \thmref{thm:VLF_Gaussian}, it holds that 
    \begin{align}
        \log M^*_{\mr{U}}(N, \epsilon, P) &\geq \frac{N C(S)}{1-\epsilon} - \left(\frac{C(S)}{C_1(S)} + \frac{1}{2} \right) \log N  \notag \\
        &\quad - \log \log N + o(\log \log N) \label{eq:GaussianUVLF}
    \end{align}
\end{theorem}

\begin{IEEEproof}
    The proof of \thmref{thm:UVLF_Gaussian} relies on a large deviations bound for the empirical correlation coefficient of two independent Gaussian distributed sequences. See \secref{sec:proof_UVLF_Gaussian} for the proof details. 
\end{IEEEproof}

\thmref{thm:UVLF_Gaussian} refines the achievability result in \cite[Th.~4]{lomnitzfeder} to the second-order term for the Gaussian channel.
To prove \thmref{thm:UVLF_Gaussian}, we replace the empirical mutual information $n I(\hat{P}_{X^n}, \hat{P}_{Y^n|X^n})$ used in the DMC case with the universal metric
\begin{align}
    \imath_{\mr{U}}(X^n; Y^n) \triangleq - \frac{n}{2} \log (1 - \hat{\rho}_{X^n Y^n}^2), 
\end{align}
where $\hat{\rho}_{X^n Y^n}$ is the empirical correlation coefficient of the zero-mean pairs $(X^n, Y^n)$ defined as
\begin{align}
    \hat{\rho}_{X^n Y^n} \triangleq \frac{\frac{1}{n} \sum_{i = 1}^n X_i Y_i}{\sqrt{\frac{1}{n} \sum_{i = 1}^n X_i^2} \sqrt{\frac{1}{n} \sum_{i = 1}^n Y_i^2}}.
\end{align}
Recall that for zero-mean, jointly Gaussian $(X, Y)$, the mutual information $I(P_X, P_{Y|X})$ is given by
\begin{align}
    I(P_X, P_{Y|X}) = - \frac{1}{2} \log (1 - \rho_{XY}^2),
\end{align}
where $\rho_{XY}$ is the correlation coefficient between $X$ and $Y$. The universal metric $ \imath_{\mr{U}}(X^n; Y^n)$ can be viewed as the empirical mutual information for the Gaussian channel in the sense that
\begin{align}
    \imath_{\mr{U}}(X^n; Y^n) = n I(\hat{P}_{X^n}^{\mr{ML}}, \hat{P}_{Y^n|X^n}^{\mr{ML}}),
\end{align}
where $(\hat{P}_{X^n}^{\mr{ML}}, \hat{P}_{Y^n|X^n}^{\mr{ML}})$ is the maximum likelihood estimator of $(P_X, P_{Y|X})$ within the family of jointly-Gaussian distributions. 

Under the Gaussian input distribution $P_X = \mathcal{N}(0, P)$, the output distribution is  $P_Y = \mathcal{N}(0, P + \sigma_0^2)$. Therefore, for the Gaussian channel, there is a one-to-one correspondence between the output distribution $P_Y$ and the channel transition kernel $P_{Y|X}$. This means that for UVLF codes over the Gaussian channel, we can use the output sequence from the first communication phase, $Y^{\tau^{(1)}}$ (without relying on the transmitted symbols $X^{\tau^{(1)}}$), to obtain an estimate $\tilde{P}_{Y|X}$, which is used in the HT phase of our Yamamoto--Itoh scheme. Note that this is not possible, e.g., for a BSC family since the output distribution $P_Y = \mr{Bernoulli}(1/2)$ is the same for all channels in the BSC family.

\section{Proofs of Theorems~\ref{thm:nonasymp} and \ref{thm:VLF}} \label{sec:proofVLF}
\subsection{Proof of \thmref{thm:nonasymp}} \label{sec:proof_nonasymp}

\subsubsection{Coding scheme} \label{sec:codingscheme}
    Let $P_X$ be a capacity-achieving input distribution, i.e., $C = I(P_X, P_{Y|X})$. As in \cite{polyanskiy2011feedback, yavas2023VLF}, we define the common randomness random variable $U$ as
    \begin{align}
        \mc{U} &\triangleq \underbrace{\mc{X}^{\infty} \times \dots \times \mc{X}^{\infty}}_{M \text{ times}} \label{eq:U1} \\
        P_U &\triangleq \underbrace{P_{X}^{\infty} \times \dots \times P_{X}^{\infty}}_{M \text{ times}}. \label{eq:U2}
    \end{align}
    The realization of $U$ defines $M$ i.i.d. infinite-length codewords from the distribution $P_X^{\infty}$.\footnote{In \cite[Th.~19]{polyanskiy2011feedback}, it is proved that $|\mc{U}|$ can be reduced to 3, implying that a code that is time-sharing of at most 3 deterministic codes can be found.} Let the generated \emph{random} codewords be $\mb{c}(1), \dots, \mb{c}(M)$. We denote the first $n$ symbols of the codeword $\mb{c}(m)$ by $\mb{c}^n(m) \triangleq (\mb{c}_1(m), \dots, \mb{c}_n(m))$. The $i$-th symbol received during communication phases (one of C1 and C2) is denoted by $Y_i$; the $i$-th symbol received during the HT phase is denoted by $\tilde{Y}_i$.

    \textbf{C1 phase:} Without loss of generality, assume that $W = 1$ is the transmitted message. Therefore, for any $n \in \mathbb{N}$,
    \begin{align}
        &P_{\mb{c}^n(1) \dots \mb{c}^n(M) Y^n}(x^n(1), \dots, x^n(M), y^n) \notag \\
        &= \prod_{i = 1}^n \left(\prod_{m = 1}^M  P_X(x_i(m)) \right) P_{Y|X}(y_i | x_i(1)), \label{eq:codedist}
    \end{align}
    where $x^n(m) = (x_1(m), \dots, x_n(m)) \in \mc{X}^n$.
    At time $n$, the transmitter transmits the $n$-th symbol $\mb{c}_n(1)$ of the codeword $\mb{c}(1)$. Let $\gamma_1,\gamma_2 \in \mathbb{R}$ be some thresholds that satisfy $\gamma_2 > \gamma_1$. For $i \in [2]$, we define the stopping times
    \begin{IEEEeqnarray}{rCl}
        \tau_m^{(i)} &\triangleq& \inf\{n \geq 1 \colon \imath(\mb{c}^n(m); Y^n) > \gamma_i\} \label{eq:taum} \\
        \tau^{(i)} &\triangleq& \min \limits_{m \in [M]} \tau_m^{(i)}, \IEEEeqnarraynumspace\label{eq:tau1def}
    \end{IEEEeqnarray}
    and the receiver's estimates
     \begin{IEEEeqnarray}{rCl}
        \hat{W}^{(i)} \triangleq \min\{ m \in [M] \colon \imath(\mb{c}^{\tau^{(i)}}(m); Y^{\tau^{(i)}}) > \gamma_i\}. \label{eq:Wi}
    \end{IEEEeqnarray}
    Through feedback, the transmitter learns whether $\tau^{(1)}$ is reached at each time during the C1 phase. This type of feedback signal that does not alter the transmitted symbol beyond telling the transmitter when to stop transmitting is called \emph{stop-feedback}. At time $\tau^{(1)}$, $\hat{W}^{(1)}$ is fed back to the transmitter for the transmitter to accept or reject $\hat{W}^{(1)}$. 

    \textbf{Hypothesis Test (HT) phase}: 
    If $\hat{W}^{(1)} = 1$, then the transmitter transmits the sequence of $(x_{\mr{A}}, x_{\mr{A}}, \dots)$; otherwise, it sends $(x_{\mr{R}}, x_{\mr{R}}, \dots)$. The receiver constructs the sequential hypothesis test
    \begin{align}
        &H_{\mr{A}} \colon \tilde{Y} \sim P_{Y|X = x_{\mr{A}}} \\
        &H_{\mr{R}} \colon \tilde{Y} \sim P_{Y|X = x_{\mr{R}}}
    \end{align}
    and Wald's sequential probability ratio test (SPRT) 
    \begin{IEEEeqnarray}{rCl}
        \tau^{\mr{HT}} \triangleq \inf\left\{n \geq 1 \colon \sum_{i = 1}^n \log \frac{P_{Y|X = x_{\mr{A}}}(\tilde{Y}_{i})}{P_{Y|X = x_{\mr{R}}}(\tilde{Y}_{i})} \notin [-a_{\mr{R}}, a_{\mr{A}}]\right\} \IEEEeqnarraynumspace\label{eq:tauHT}
    \end{IEEEeqnarray}
    where $-a_{\mr{R}}$ and $a_{\mr{A}}$ are thresholds of the SPRT. Here, $H_{\mr{A}}$ and $H_{\mr{R}}$ correspond to hypothesis to accept and to reject the initial estimate $\hat{W}^{(1)}$, respectively.
    
    If $\sum_{i = 1}^{\tau^{\mr{HT}}} \log \frac{P_{Y|X = x_{\mr{A}}}(\tilde{Y}_{i})}{P_{Y|X = x_{\mr{R}}}(\tilde{Y}_{i})} > a_{\mr{A}}$, then $H_{\mr{A}}$ is declared at time $\tau^{(1)} + \tau^{\mr{HT}}$ by the receiver, and the initial estimate $\hat{W}^{(1)}$ is accepted as $\hat{W}$. If $\sum_{i = 1}^{\tau^{\mr{HT}}} \log \frac{P_{Y|X = x_{\mr{A}}}(\tilde{Y}_{i})}{P_{Y|X = x_{\mr{R}}}(\tilde{Y}_{i})} < - a_{\mr{R}}$, then $H_{\mr{R}}$ is declared, and the communication enters the C2 phase. The transmitter learns the receiver's decision at the end of the HT phase through feedback.

    \textbf{C2 phase:} The transmitter continues to transmit symbols from $\mb{c}(1)$ starting from the index $\tau^{(1)} + 1$ and ending at the index $\tau^{(2)}$, i.e., the symbols $\mb{c}_{\tau^{(1)} + 1}(1), \mb{c}_{\tau^{(1)} + 2}(1), \dots, \mb{c}_{\tau^{(2)}}(1)$ are transmitted. These symbols are transmitted starting from time (of the communication epoch) $\tau^{(1)} + \tau^{\mr{HT}} + 1$ and ending at time $\tau^{(2)} + \tau^{\mr{HT}}$.
    At time $\tau^{(2)} + \tau^{\mr{HT}}$, the receiver decodes to the estimate $\hat{W}^{(2)}$. The coding scheme is summarized in Table~\ref{tab}, below.
    In the proof of \thmref{thm:nonasymp}, we use the bound
    \begin{align}
        \Prob{\tau_2^{(i)} < \infty} \leq \exp\{-\gamma_i\} \label{eq:taumibound}
    \end{align}
    from \cite[eq.~(118)]{polyanskiy2011feedback} to bound the error probability terms associated with the communication phases. The error probability terms associated with the SPRT are bounded using \cite[Th.~3.1]{woodroofebook}, which is essentially equivalent to \eqref{eq:taumibound}. To bound the average decoding time of the code, we use \lemref{lem:tau}, below.

\begin{table*}[htbp!] 
\caption{The summary of our modified Yamamoto--Itoh scheme}
\centering
\label{tab}
\begin{tabular}{cccc}
 Phases& \textbf{Communication 1 (C1)} & \textbf{Confirmation (HT)} & \textbf{Communication 2 (C2)} \\
\hline 
Coding scheme    & variable-length i.i.d. random coding    &  SPRT   & variable-length i.i.d. random coding    \\
Decoding metric & information density & log-likelihood ratio & information density \\
Random length    & $\tau^{(1)}$  & $\tau^{\mr{HT}}$   & $\tau^{(2)} - \tau^{(1)}$   \\
Feedback during the phase & $\{\text{continue}, \text{end phase}\}$    &$\{\text{continue}, \text{end phase}\}$  & $\{\text{continue}, \text{end phase}\}$     \\
Feedback at the end of the phase  & $\lceil \log_2 M \rceil$ bits for $\hat{W}^{(1)}$  & $\{\text{accept } \hat{W}^{(1)}, \text{reject } \hat{W}^{(1)}\}$ & \ding{55}  \\
Condition to enter    & \ding{55}    & \ding{55}  & SPRT outputs ``reject''   
\end{tabular}
\end{table*}

    \subsubsection{Error probability analysis} Define the error events
    \begin{align}
        \mc{E}^{(i)} &\triangleq \{ \hat{W}^{(i)} \neq 1\}, \quad i = 1, 2 \\
        \mc{E}_{\mr{A} \to \mr{R}} &\triangleq \{ \mr{H}_{\mr{R}} \text { is declared given } \mr{H}_{\mr{A}} \} \\
        \mc{E}_{\mr{R} \to \mr{A}} &\triangleq \{ \mr{H}_{\mr{A}} \text { is declared given } \mr{H}_{\mr{R}} \} \\
        \mc{E}_{\mr{C2}} &\triangleq \{\text{C2 phase is entered}\}.
    \end{align}
    Then the error probability of the above scheme is bounded as
    \begin{align}
        \Prob{\hat{W} \neq 1} &\leq \Prob{(\mc{E}^{(1)} \bigcap \mc{E}_{\mr{R} \to \mr{A}}) \bigcup \mc{E}^{(2)}} \\
        &\leq \Prob{\mc{E}^{(1)}} \Prob{\mc{E}_{\mr{R} \to \mr{A}}} + \Prob{\mc{E}^{(2)}}, \label{eq:unionb}
    \end{align}
    where \eqref{eq:unionb} follows from the union bound and the independence of the events $\mc{E}^{(1)}$ and $\mc{E}_{\mr{R} \to \mr{A}}$.
    From \cite[Th.~3.1]{woodroofebook}, the type-I and type-II error probabilities of the sequential hypothesis test are bounded as
    \begin{align}
        \Prob{\mc{E}_{\mr{A} \to \mr{R}}} &\leq \exp\{- a_{\mr{R}}\} \label{eq:EAR}\\
        \Prob{\mc{E}_{\mr{R} \to \mr{A}}} &\leq \exp\{- a_{\mr{A}}\}. \label{eq:ERA}
    \end{align}
    The probabilities $\Prob{\mc{E}^{(i)}}$, $i = 1, 2$, are bounded following \cite[Proof of Th.~2]{polyanskiy2011feedback} as
    \begin{align}
         \Prob{\mc{E}^{(i)}} &\leq \Prob{\tau^{(i)}_1 = \infty} + \Prob{\bigcup_{m = 2}^M \{\tau^{(i)}_m < \infty\}} \\
         &\leq (M-1) \exp\{-\gamma_i\}, \quad i = 1, 2. \label{eq:E12}
    \end{align}
    Combining \eqref{eq:unionb}, \eqref{eq:ERA}, and \eqref{eq:E12}, we get
    \begin{align}
        \Prob{\hat{W} \neq 1} &\leq (M-1) \left(\exp\{-(\gamma_1 + a_{\mr{A}})\} + \exp\{-\gamma_2\} \right). 
        \label{eq:errorbound}
    \end{align}
    \subsubsection{Average decoding time analysis}
    We use the following result from the renewal theory literature, which bounds the expected value of the stopping time associated with a random walk.
    \begin{lemma}[\hspace{-.05em}{\cite[Th.~1]{lorden1970}, \cite[Ch. 3, Th.~9.2--9.3, Th.~10.7]{gutbook}}]\label{lem:tau}
        Let $X, X_1, X_2, \dots$ be i.i.d. random variables with $\E{X} = \mu > 0$ and $\E{(X^+)^2} < \infty$. Let $S_n = \sum_{i = 1}^n X_i$ and $\tau = \inf\{n \geq 1 \colon S_n > a\}$. Then, for any $a > 0$,
        \begin{align}
            \E{\tau} \leq \frac{1}{\mu} \left(a + b(P_X) \right).  
        \end{align}
        Let 
        \begin{align}
            \tau_+ &= \inf\{n \geq 1 \colon S_n > 0\} \label{eq:tauplus} \\
            \rho &= \frac{\E{S^2_{\tau_+}}}{2 \E{S_{\tau_+}}}. \label{eq:rho}
        \end{align}
        As $a \to \infty$, if $X$ is non-arithmetic and the above conditions are satisfied, then 
        \begin{align}
            \E{\tau} = \frac{1}{\mu}\left(a + \rho \right) + o(1), \label{eq:taurenewal}
        \end{align}
        and if $X$ is arithmetic with a span $h$ and $a = j h$, $j \in \mathbb{Z}$, $j \to \infty$, then
        \begin{align}
            \E{\tau} = \frac{1}{\mu}\left(a + \rho + \frac{h}{2} \right) + o(1).
        \end{align}
    It holds that
    \begin{align}
        \frac{\E{S^2_{\tau_+}}}{2 \E{S_{\tau_+}}} = \frac{\E{X^2}}{2 \mu} - \sum_{k = 1}^{\infty} \frac{1}{k} \E{S_k^-}.
    \end{align}

    \end{lemma}
    
    We bound the probability that the C2 phase is used as
    \begin{align}
        \Prob{\mc{E}_{\mr{C2}}} &= \Prob{(\mc{E}^{(1)} \cap \mc{E}_{\mr{R} \to \mr{A}}^{\mr{c}}) \cup ((\mc{E}^{(1)})^{\mr{c}} \cap \mc{E}_{\mr{A} \to \mr{R}})} \\
        &\leq \Prob{\mc{E}^{(1)}} + \Prob{\mc{E}_{\mr{A} \to \mr{R}}} \\
        &\leq (M-1) \exp\{-\gamma_1\} + \exp\{-a_{\mr{R}}\}. \label{eq:E2}
    \end{align}

    Recall the stopping times defined in \eqref{eq:taum}--\eqref{eq:tau1def}. Obviously, it holds for the stopping times defined in \eqref{eq:taum}--\eqref{eq:tau1def} that $\tau^{(i)}_1 \leq \tau^{(i)}$ for $i = 1, 2$. By our code design, $\tau = \tau^{(1)} + \tau^{\mr{HT}}$ if the event $\mc{E}_{\mr{C2}}$ does not occur and $\tau = \tau^{(2)} + \tau^{\mr{HT}}$ if $\mc{E}_{\mr{C2}}$ occurs. Therefore, we bound the average stopping time as
    \begin{align}
        \E{\tau} \leq \E{\tau^{(1)}_1} + \Prob{\mc{E}_{\mr{C2}}} \E{\tau^{(2)}_1 - \tau^{(1)}_1} + \E{\tau^{\mr{HT}}}. \label{eq:taubound}
    \end{align}
    Applying \lemref{lem:tau}, we bound each of the expectations in \eqref{eq:taubound} as 
    \begin{align}
        \E{\tau_{\mr{HT}} | H_{\mr{A}}} &\leq \frac{a_{\mr{A}} + b_{\mr{A}}}{D(P_{Y|X = x_{\mr{A}}} \| P_{Y|X = x_{\mr{R}}})} \label{eq:THTA} \\
        \E{\tau_{\mr{HT}} | H_{\mr{R}}} &\leq \frac{a_{\mr{R}} + b_{\mr{R}}}{D(P_{Y|X = x_{\mr{R}}} \| P_{Y|X = x_{\mr{A}}})} \label{eq:THTR} \\
        \E{\tau_{\mr{HT}}} &\leq \E{\tau_{\mr{HT}} | H_{\mr{A}}}  + \Prob{H_\mr{R}} \E{\tau_{\mr{HT}} | H_{\mr{R}}} \\
        &\leq \E{\tau_{\mr{HT}} | H_{\mr{A}}} \notag \\
        &\quad + (M-1) \exp\{-\gamma_1\} \E{\tau_{\mr{HT}} | H_{\mr{R}}} \label{eq:tauHTbound} \\
        \E{\tau^{(1)}_1} &\leq \frac{\gamma_1 + b}{C} \label{eq:tau1bound} \\
        \E{\tau^{(2)}_1 - \tau^{(1)}_1} &\leq \frac{\gamma_2 - \gamma_1 + b}{C}, \label{eq:tau1tau2}
    \end{align}
    where $b = b(P_{\imath(X; Y)})$,  $b_{\mr{A}} = b \left(P_{\log \frac{P_{Y|X = x_{\mr{A}}}(Y_{\mr{A}})}{P_{Y|X = x_{\mr{R}}}(Y_{\mr{A}})}}\right)$, and $b_{\mr{R}} = b\left(P_{\log \frac{P_{Y|X = x_{\mr{R}}}(Y_{\mr{R}})}{P_{Y|X = x_{\mr{A}}}(Y_{\mr{R}})}}\right)$, and and $Y_\mr{D}$ is distributed according to $P_{Y|X = x_{\mr{D}}}$ for $\mr{D} \in \{\mr{A}, \mr{R}\}$. In \eqref{eq:tauHTbound}, we use $\Prob{H_{\mr{R}}} = \Prob{\mc{E}^{(1)}}$ and the bound in \eqref{eq:E12}.

    Finally, combining \eqref{eq:E2}, \eqref{eq:taubound}, and \eqref{eq:tauHTbound}--\eqref{eq:tau1tau2} gives
    \begin{align}
        \E{\tau} &\leq \frac{\gamma_1 + b}{C} \notag \\
        &+ ((M-1) \exp\{-\gamma_1\} + \exp\{-a_{\mr{R}}\}) \frac{\gamma_2 - \gamma_1 + b}{C} \notag \\
        &+\frac{a_{\mr{A}} + b_{\mr{A}}}{D(P_{Y|X = x_{\mr{A}}} \| P_{Y|X = x_{\mr{R}}})} \notag\\
        &+ (M-1) \exp\{-\gamma_1\} \frac{a_{\mr{R}} + b_{\mr{R}}}{D(P_{Y|X = x_{\mr{R}}} \| P_{Y|X = x_{\mr{A}}})}. \label{eq:Nbound}
    \end{align}
    From the above analysis, there exists an $(N', M, \epsilon')$-VLF code where $\epsilon'$ and $N'$ are given as the right hand sides of \eqref{eq:errorbound} and \eqref{eq:Nbound}, respectively. We use the stop-at-time-zero strategy described in \cite{polyanskiy2011feedback}, where with probability $1-\epsilon_0$, the code above is used, and with probability $\epsilon_0$, we use a simple code that stops at time zero and decodes to an arbitrary message. Let $N$ and $\epsilon$ be the average decoding time and the average error probability of the described code obtained by this time-sharing strategy. We have
    \begin{align}
        N &\leq (1-\epsilon_0) N' \\
        \epsilon &\leq \epsilon_0 + (1-\epsilon_0) \epsilon',
    \end{align}
    which completes the proof.

\subsection{Proof of \thmref{thm:VLF}} \label{sec:proofVLF_2}
We prove \thmref{thm:VLF} by carefully choosing the free parameters $\gamma_1, \gamma_2, a_{\mr{A}}, a_{\mr{R}}$, and $\epsilon_0$ in \thmref{thm:nonasymp}. Let
\begin{align}
    N_1 = \frac{\gamma_1 + b}{C}, \label{eq:N1gamma}
\end{align}
which is an upper bound on the expected length of the C1 phase. We express all other parameters in terms of $N_1$. 
We set
\begin{align}
    \gamma_1 &= \log M + \log \log N_1 \label{eq:gammaM} \\
    \gamma_2 &= \log M + \log N_1 \label{eq:gamma2M}\\
    a_\mr{A} &= a_{\mr{R}} = \log N_1. 
\end{align}
Then, by \eqref{eq:Nprime_bound}, we have as $N_1 \to \infty$
\begin{align}
    N' &= N_1 + \frac{\log N_1}{C_1} + O(1) \label{eq:NprimeN1}\\
    \epsilon' &\leq \frac{1}{N_1} \left(1 + \frac{1}{\log N_1} \right). \label{eq:epsprime_bound1}
\end{align}
We set
\begin{align}
    \epsilon_0 = \frac{\epsilon - \frac{1}{N_1} \left(1 + \frac{1}{\log N_1} \right)}{1 - \frac{1}{N_1} \left(1 + \frac{1}{\log N_1} \right)} = \epsilon - \Omega \left(\frac{1}{N_1} \left( 1 + \frac{1}{\log N_1} \right)\right). \label{eq:eps0} 
\end{align}
From \eqref{eq:N1gamma} and \eqref{eq:gammaM}, we get
\begin{align}
    \log M = N_1 C - \log \log N_1 + O(1). \label{eq:logM1}
\end{align}
From \eqref{eq:NprimeN1} and \eqref{eq:logM1}, we get
\begin{align}
    \log M = N' C - \frac{C}{C_1} \log N' - \log \log N' + O(1). \label{eq:logMNprime}
\end{align}
Finally, from \eqref{eq:epsbound} and \eqref{eq:eps0}, the error probability of the code is bounded by $\epsilon$,
and the average decoding time of the code is bounded by $(1-\epsilon_0) N'$. Therefore, by \eqref{eq:NprimeN1} and \eqref{eq:logMNprime}, there exists an $(N, M, \epsilon)$-VLF code with 
\begin{align}
     \log M = \frac{N C}{1-\epsilon} - \frac{C}{C_1} \log N - \log \log N + O(1). \label{eq:lastlogM}
\end{align}

\section{Proof of \thmref{thm:UVLF}} \label{sec:proofUVLF}
\subsection{Supporting Lemmas}
We first present two supporting lemmas that play key roles to prove  \thmref{thm:UVLF}. The first result, below, bounds the tail probability of the empirical mutual information for independent $\bar{X}^n$ and $Y^n$.
\begin{lemma}
\label{lem:empMI}
Let $(\bar{X}^n, Y^n) \sim P_X^n P_Y^n$ for some $P_X \in \mc{P}(\mc{X})$ and $P_Y \in \mc{P}(\mc{Y})$, and let $\gamma$ be a positive constant. Assume that $P_X(x) > 0$ and $P_Y(y) > 0$ for all $(x, y) \in (\mc{X} \times \mc{Y})$. Then, there exists $n_0 \in \mathbb{N}$ such that for all $n \geq n_0$
\begin{align}
    &\Prob{n I(\hat{P}_{\bar{X}^n}, P_{Y^n|\bar{X}^n}) \geq \gamma} \leq K_1 (n + 1)^d \exp\{-\gamma\} \label{eq:I1} \\
    &d = \min \left\{{\frac{|\mc{X}| |\mc{Y}| - 2}{2}},{\left(|\mc{X}|-\frac{3}{2}\right) \left(|\mc{Y}|-\frac{3}{2}\right) - \frac{1}{4}} \right\}, \label{eq:I2}
\end{align}
where $K_1$ is a positive constant depending only on $|\mc{X}|$ and $|\mc{Y}|$. 
\end{lemma}
\begin{IEEEproof}
See \appref{app:proof_EMI}.
\end{IEEEproof}

The second result, which is from the nonlinear renewal theory literature, bounds the expected stopping time associated with a function of an i.i.d. sum. This result is the nonlinear version of \lemref{lem:tau} and is used to bound the expected stopping times associated with the empirical mutual information. 
\begin{lemma}[\hspace{-.05em}{\cite[Th.~4.5]{woodroofebook}}]\label{lem:nonlinear}
    Let $g \colon \mathbb{R}^k \to \mathbb{R}$ be a twice differentiable continuous function. Let $Y, Y_1, Y_2, \dots \in \mathbb{R}^k$ be i.i.d. random vectors. Let $\mu = g(\E{Y}) > 0$. Let $\gamma > 0$. Define 
    \begin{align}
        Z_n &= n g\left( \frac{1}{n} \sum_{i = 1}^n Y_i \right) \\
        \tau &= \inf\{ n \geq 1 \colon Z_n > a\}. 
    \end{align}
    Then, if $\mu + \nabla g(\E{Y})^\top (Y - \E{Y})$ is non-arithmetic, as $a \to \infty$,
    \begin{align}
        \E{\tau} = \frac{1}{\mu} \left(a + \rho - \frac{1}{2}\mr{tr}(\mr{Cov}(Y) \nabla^2 g(\E{Y})) \right) + o(1),
    \end{align}
    where $\rho$ is defined in \eqref{eq:rho} with $S_n = \sum_{i = 1}^n X_i$ replaced with $n \mu + \sum_{i = 1}^n \nabla g(\E{Y})^\top (Y_i - \E{Y})$. If $\mu + \nabla g(\E{Y})^\top (Y - \E{Y})$ is arithmetic with a span $h > 0$, then for $a = j h$, $j \in \mathbb{Z}$, $j \to \infty$,
        \begin{align}
        \E{\tau} = \frac{1}{\mu} \left(a + \rho + \frac{h}{2} - \frac{1}{2}\mr{tr}(\mr{Cov}(Y) \nabla^2 g(\E{Y})) \right) + o(1).
    \end{align}
\end{lemma}

\lemref{lem:nonlinear} is a special case of the nonlinear form $Z_n = \sum_{i = 1}^n X_i + \xi_n$, where $X_1, X_2, \dots$ are i.i.d. random variables, $\xi_n$'s are slowly changing random variables, which is specified in \cite[eq.~(4.10)--(4.16)]{woodroofebook}, and $(X_1, \xi_1), \dots, (X_n, \xi_n)$ are independent of $X_k, k > n$. From the Taylor series expansion of $Z_n$ around $n g(\E{Y})$, we get
\begin{align}
    &Z_n = n \mu + \sum_{i = 1}^n \nabla g(\E{Y})^\top (Y_i - \E{Y}) \notag \\
    &+  \frac{1}{2} W_n ^\top \nabla^2 g(\E{Y}) W_n  + o(1) \label{eq:Zn} \\
    &W_n = \frac{1}{\sqrt{n}} \sum_{i = 1}^n (Y_i - \E{Y}).
\end{align}
Therefore, in \lemref{lem:nonlinear}, $\mu + \nabla g(\E{Y})^\top (Y_i - \E{Y})$ plays the role of $X_i$ and $\frac{1}{2} W_n ^\top \nabla^2 g(\E{Y}) W_n  + o(1)$ plays the role of $\xi_n$. In \cite[Example~4.1]{woodroofebook}, it is shown that $W_n$ satisfies the slowly-changing conditions. By the central limit theorem, $W_n$ approaches the Gaussian vector $\mathcal{N}(0, \mr{Cov}(Y))$ in distribution, and the third-term in \eqref{eq:Zn} approaches the sum of $k$ independent $\chi^2(1)$ random variables, weighted with $\frac{1}{2}$ times the eigenvalues of the matrix $\mr{Cov}(Y) \nabla^2 g(\E{Y})$.

\subsection{Universal Coding Scheme} \label{sec:univcode}

The proposed code is an UVLF code that employs stop-feedback. Our code universalizes Polyanskiy \emph{et al.}'s code in \cite{polyanskiy2011feedback} by employing the empirical mutual information as its decoding metric instead of the information density. Compared to the Yamamoto--Itoh scheme described in \secref{sec:main}, the proposed UVLF code essentially accepts the estimate at the end of the C1 phase as its final decision.

We generate $M$ i.i.d.\ codewords $\mb{c}(1), \dots, \mb{c}(M)$ each from $P_X^{\infty}$ as in \eqref{eq:U1}--\eqref{eq:U2}. 
Let
\begin{align}
        \tau_m &= \inf\{n \geq 1 \colon n I(\hat{P}_{\mb{c}^n(m)}, \hat{P}_{Y^n|\mb{c}^n(m)}) > \gamma \}, \label{eq:tau_UVLF}
\end{align}
where
\begin{align}
    \gamma &= \log M + (d + 1) \log n_1 + \delta \log \log n_1 \label{eq:gammastop} \\
    n_1 &\triangleq \left \lfloor \frac{\log M}{\log \min\{|\mc{X}|, |\mc{Y}|\}} \right \rfloor, 
\end{align}
$d$ is the constant given in $\eqref{eq:I2}$, and $\delta > 0$ is an arbitrarily small constant. Note that $n_1$ is a lower bound on $\tau$ since $\gamma \geq \log M$ and $I(P_X, P_{Y|X}) \leq \log \min \{|\mc{X}|, |\mc{Y}|\}$. It holds 
that $n_1 = \Theta(N)$ for every DMC with $C > 0$.

The decoder stops at time $\tau \triangleq \min_{m \in [M]} \tau_m$ and decodes to a message $\hat{W}$ with $\tau_{\hat{W}} = \tau$. Through stop-feedback, the transmitter learns when $\tau$ is reached.

\subsection{Analysis} \label{sec:UVLF_ana}
We here explain the differences from the proof of \thmref{thm:VLF}. 
\begin{enumerate}
    \item Let $\mc{E}$ be the error probability of the code (the stop-at-time-zero component is not considered yet), i.e., 
    \begin{align}
        \mc{E} \triangleq \{\hat{W} \neq 1\}.
    \end{align}
    
    Let $n_2 \triangleq c_2 \frac{\log M}{C}$, where $c_2 > 1$ be a sufficiently large constant, which gives $n_2 = \Theta(n_1)$. We bound the probability $\epsilon' \triangleq \Prob{\mc{E}}$ as
    \begin{align}
        \epsilon' &\leq \Prob{\tau_1 \geq n_2} + \Prob{\bigcup_{m = 2}^M \{\tau_m < n_2\}} \label{eq:E1first} \\
        &\leq \Prob{n_2 I(\hat{P}_{\mb{c}^{n_2}(1) }, \hat{P}_{Y^{n_2}|\mb{c}^{n_2}(1)}) \leq \gamma} \notag \\
        &\quad + (M-1) (n_2 - n_1) \notag \\
        &\quad \max \limits_{n \in [n_1, n_2 - 1]} \Prob{n I(\hat{P}_{\mb{c}^n(2)}, \hat{P}_{Y^n|\mb{c}^n(2)}) > \gamma} \label{eq:step1}\\
        &\leq n_2^{|\mc{X}||\mc{Y}|} \exp\{-c_3 n_2\} + K_1 M n_2^{d} \exp\{-\gamma\}, \label{eq:step2} \\
        &\leq \exp\{-\Omega(n_1)\} + \frac{1}{n_1} \frac{K_1 (\frac{n_2}{n_1})^d }{(\log n_1)^{\delta}} \\
    &\leq \frac{1}{n_1} \label{eq:epsprimestop}
\end{align}
for $n_1$ large enough. Here,
    $c_3$ is a positive constant independent of $n_2$, and $K_1$ and $d$ are given in \eqref{eq:I1}--\eqref{eq:I2}. Inequality  \eqref{eq:step1} follows from the definition of $\tau$ and the union bound across time and messages. The first term in \eqref{eq:step2} follows from the standard method of types (see e.g., \cite[Lemma~3]{tchamkerten2006variable}) provided that $c_2 > \frac{\gamma}{\log M}$, which holds due to \eqref{eq:gammastop} and $c_2 > 1$. The second term in \eqref{eq:step2} follows from \lemref{lem:empMI}. 

    \item We bound the expected value of the stopping time $\tau$ using \lemref{lem:nonlinear} instead of \lemref{lem:tau}. To do this, we write the empirical mutual information as
    \begin{align}
        &n I(\hat{P}_{X^n}, \hat{P}_{Y^n|X^n}) = n I\left(\sum_{i = 1}^n V_i\right) \\
        &= \sum_{i = 1}^n \imath(X_i; Y_i) + \frac{1}{2} W_n^\top \nabla^2 I(P_X, P_{Y|X}) W_n + o(1)\\
        &W_n = \frac{1}{\sqrt{n}} \sum_{i = 1}^n (V_i - \E{V_i}),
    \end{align}
    where $V_i \in \mathbb{R}^{|\mc{X}||\mc{Y}|}$, $i = 1, \dots, n$, are independent and have multinomial distribution with parameters $(n, P_{XY})$. Hence, $\imath(X^n; Y^n)$ equals the first-order Taylor approximation to the empirical mutual information $n I(\hat{P}_{X^n}, \hat{P}_{Y^n |X^n})$. Applying \lemref{lem:nonlinear} to $\E{\tau_1}$, similarly to \eqref{eq:tau1bound}--\eqref{eq:tau1tau2}, we get
    \begin{align}
        N' \leq \E{\tau_1} &= \frac{\gamma}{C} + O(1) \label{eq:newtau1}
    \end{align}
    Notice that the bound in \eqref{eq:newtau1} is asymptotically the same as the bound in \eqref{eq:tau1bound} except that the value of the constant $O(1)$ term differs. 
  
    \item Combining \eqref{eq:gammastop}, \eqref{eq:epsprimestop}, and \eqref{eq:newtau1}, we show that there exists an $(N', M, \epsilon')$-UVLF code with
\begin{align}
    \log M = N' C - (d + 1) \log N' + o(\log \log N). \label{eq:logMUVLF}
\end{align}
We set the stopping probability at time zero as
\begin{align}
    \epsilon_0 = \frac{\epsilon - \frac{1}{n_1}}{1 - \frac{1}{n_1}}, \label{eq:neweps0}
\end{align}
which ensures that the overall error probability of our code is bounded by $\epsilon$.
Following steps similar to those in \eqref{eq:eps0}--\eqref{eq:lastlogM} with the modifications in \eqref{eq:logMUVLF}--\eqref{eq:neweps0} and $n_1 =\Theta(N_1)$, we complete the proof of \eqref{eq:UVLF}.
    
\end{enumerate}

\subsection{Universal Coding Scheme for a BSC Family and Its Analysis}
The coding scheme is identical to that in \secref{sec:univcode} except that the stopping time in \eqref{eq:tau_UVLF} is replaced with
\begin{align}
    \tau_{m} \triangleq \inf\{ n \geq 1 \colon n(\log 2 - H(\hat{P}_{Z^n(m)})) > \gamma\}, \label{eq:BSCtau}
\end{align}
where $Z_i(m) = 1\{Y_i \neq \mb{c}_i(m)\}$. 
 Define
\begin{align}
    \hat{p}_n(m) \triangleq \frac{1}{n} \sum_{i = 1}^n Z_i(m).
\end{align}
Hence, $H(\hat{P}_{Z^n(m)}) = h_\mr{b}(\hat{p}_n(m))$ is the binary entropy function of the empirical flip probability from the sub-codeword $\mb{c}^n(m)$ to the output sequence $Y^n$. 

We bound the probability $\Prob{\tau_2 < n_2}$ differently than \eqref{eq:step1}--\eqref{eq:step2}. The information density under the BSC($p$) equals
\begin{align}
    \imath(\mb{c}^n(m); Y^n) = n \left(\log(2(1-p)) -  \hat{p}_n(m) \log \frac{1-p}{p}\right).
\end{align}
Therefore, both $\tau_1$ and $\imath(\mb{c}^n(1); Y^n)$ depends on $(\mb{c}^n(1), Y^n)$ only through the empirical flip probability $\hat{p}_n(1)$. This means that $\tau_1$ is a stopping time for the martingale $\{\exp\{-\imath(\mb{c}^n(1); Y^n)\}\}_{n \geq 1}$. Using this property, we apply the steps in \cite[eq.~(111)--(117)]{polyanskiy2011feedback} and get
\begin{align}
    \Prob{\tau_2 < n_2} = \E{\exp\{-\imath(\mb{c}^{\tau_1}(1); Y^{\tau_1})\} 1\{\tau_1 < n_2\}}, \label{eq:exp}  
\end{align}
which follows from a changing measure argument and Doob's optional stopping theorem.  
Define
\begin{align}
    R &\triangleq \tau_1(\log 2 - H(\hat{P}_{Z^{\tau_1}(1)})) - \gamma_i \\
    \eta_n &\triangleq n(\log 2 - H(\hat{P}_{Z^n(1)})) - \imath(\mb{c}^n(1); Y^n).
\end{align}
Here, $R \geq 0$ is the overshoot random variable corresponding to the transmitted codeword. Then, we bound the right-hand side of \eqref{eq:exp} as
\begin{align}
    &\Prob{\tau_2 < n_2} \notag \\
    &= \exp\{-\gamma\} \E{\exp\{-R + \eta_{\tau_1}\} 1\{\tau_1 < n_2\}} \\
    &\leq \exp\{-\gamma\} \E{\exp\{ \eta_{\tau_1}\} 1\{\tau_1 < n_2\}} \\
    &\leq \max_{n < n_2} \E{\exp\{\eta_n\}} \exp\{-\gamma\} \\
    &\leq K_2 \sqrt{n_2} \exp\{-\gamma\}, \label{eq:etastep}
\end{align}
where $K_2$ is a positive constant independent of $n_2$ and $\gamma$. The last step in \eqref{eq:etastep} is proved in \appref{app:eta}. 

We check that the first-order Taylor series expansion to the universal metric $n (\log 2 - H(\hat{P}_{Z^n}))$ is equal to the information density $\imath(X^n; Y^n)$, where $Z_i = 1\{Y_i \neq X_i\}$, $i \in [n]$. Therefore, the asymptotic expansion on the right-hand side of \eqref{eq:newtau1} remains to hold.

Comparing \eqref{eq:step2} with \eqref{eq:etastep}, we set $d = \frac{1}{2}$ in \eqref{eq:gammastop}, then follow the same steps as in \eqref{eq:logMUVLF}--\eqref{eq:neweps0} to complete the proof of \eqref{eq:BSCUVLF}.

\section{Proof of \thmref{thm:UVLF_Gaussian}} \label{sec:proof_UVLF_Gaussian}
The following result is a strong large deviations bound for the correlation coefficient of the two jointly-Gaussian random variables.
\begin{lemma}[\hspace{-.05em}{\cite[Th.~3.5]{truong:hal}}] \label{lem:rho}
    Let $(X^n, Y^n)$ be i.i.d. from $\mathcal{N}(0, \mathsf{\Sigma})$ and $\ms{\Sigma}_{12} = 0$, i.e., $X_i$ and $Y_i$ are independent. Let $\hat{\rho}_{X^n Y^n} = \frac{\frac{1}{n} \sum_{i = 1}^n X_i Y_i}{\sqrt{\frac{1}{n} \sum_{i = 1}^n X_i^2} \sqrt{\frac{1}{n} \sum_{i = 1}^n Y_i^2}}$ be the empirical correlation coefficient. Let $a \in (0, 1)$. Then, as $n \to \infty$,
    \begin{align}
        \Prob{\hat{\rho}_{X^n Y^n} \geq a} &= \frac{(1 - 4 \lambda_a^2)^{-\frac{1}{4}}}{\lambda_a \sigma_a \sqrt{n}} \exp\left\{\frac{n}{2} \log (1 - a^2) \right\} \notag \\
        & \quad \cdot (1 + o(1)), 
    \end{align}
    where
    \begin{align}
        \sigma_a &= \frac{1 - a^2}{\sqrt{1 + a^2}} \\
        \lambda_a &= \frac{a}{1 - a^2}.
    \end{align}
\end{lemma}
\lemref{lem:rho} replaces the role of \lemref{lem:empMI} for the Gaussian channel.

\subsection{Universal Coding Scheme for the Gaussian Channel}
The code employs the universal version of the Yamamoto--Itoh code described in \secref{sec:main}.
We here explain the differences from the coding scheme in \secref{sec:main}.

We generate $M$ i.i.d.\ codewords $\mb{c}(1), \dots, \mb{c}(M)$ each from $P_X^{\infty} = \mc{N}(0, P)^{\infty}$. The stopping times $\tau_m^{(i)}$ in \eqref{eq:taum} are replaced with
\begin{align}
    \tau_m^{(i)} = \inf \{n \geq 1 \colon \imath_{\mr{U}}(\mb{c}^n(m); Y^n) > \gamma_i\}. \label{eq:tauGaussian}
\end{align}
Recall that
\begin{align}
    \tau^{(i)} = \min_{m \in [M]} \tau_m^{(i)},
\end{align}
and $\tau^{(1)}$ is the time index for the end of the C1 phase. 
The channel $P_{Y|X}$ that is used in the HT phase in \eqref{eq:tauHT} is replaced with
\begin{align}
    \tilde{P}_{Y|X = x} = \mc{N}\left(x, \frac{1}{\tau^{(1)}} \left(\sum_{i = 1}^{\tau^{(1)}} Y_i^2\right) - P \right), \label{eq:newtilde}
\end{align}
where $Y^{\tau^{(1)}}$ is the output sequence observed in the C1 phase. Note that $\E{Y_i^2} = P + \sigma_0^2$ is a one-to-one function of $P_{Y|X}$. Hence \eqref{eq:newtilde} gives an estimate of the unknown channel $P_{Y|X}$. 

Let
\begin{align}
    n_1 = \lfloor \log M \rfloor,
\end{align}
which satisfies $n_1 = \Theta(N)$ for any $\sigma_0^2 > 0$. The stopping time \eqref{eq:tauHT} for the HT phase is replaced with
\begin{align}
        \tau^{\mr{HT}} &\triangleq \min\bigg\{\inf\bigg\{n \geq 1 \colon \sum_{i = 1}^n \log \frac{\tilde{P}_{Y|X = x_{\mr{A}}}(\tilde{Y}_{i})}{\tilde{P}_{Y|X = x_{\mr{R}}}(\tilde{Y}_{i})} \notag \\
        &\quad \notin [-a_{\mr{R}}, a_{\mr{A}}] \bigg\}, n_1 \bigg\}, \label{eq:tauHT_new}
\end{align}
where we set $x_{\mr{A}} = \sqrt{P}$ and $x_{\mr{R}} = - \sqrt{P}$.

We define the typical event $\mc{G}$ as
\begin{align}
    \mc{G} \triangleq \left \{ \left \lvert \left(\frac{1}{\tau^{(1)}} \sum_{i = 1}^{\tau^{(1)}} Y_i^2 \right) - P  - \sigma_0^2  \right \rvert \leq \sqrt{\frac{\log n_1}{n_1 }} \right \}. \label{eq:Pcond}
\end{align}
Applying the Chernoff bound to the chi-squared random variable, we have
\begin{align}
    \Prob{\mc{G}^{\mr{c}}} \leq O\left(\frac{1}{n_1^2}\right).
\end{align}

In the following, we explain the differences from the proof steps in \secref{sec:proof_nonasymp}--\ref{sec:UVLF_ana}. 
\begin{enumerate}
    \item Let $n_2 = c_2 \frac{\gamma_2}{C(S)}$, where $c_2 > 1$ is a sufficiently large constant. Recall that $\mc{E}^{(i)} = \{\hat{W}^{(i)} \neq 1\}$. We bound the probability $\Prob{\mc{E}^{(i)}}$ as
    \begin{align}
        &\Prob{\mc{E}^{(i)}} \leq \Prob{\tau^{(i)}_1 \geq n_2} + \Prob{\bigcup_{m = 2}^M \{\tau^{(i)}_m < n_2\}} \\
        &\leq \Prob{\imath_{\mr{U}}(\mb{c}^{n_2}(1); Y^{n_2}) \leq \gamma_i} \notag \\
        &+ (M-1) \sum_{n = 1}^{n_2}
        \Prob{\imath_{\mr{U}}(\mb{c}^n(2); Y^n) > \gamma_i} \label{eq:step1U}\\
        &\leq \exp\{-c_4 n_2\} + K_3 n_2^{1/2} \exp\{-\gamma_i\}, \label{eq:step2U}
    \end{align}
    where $c_4$ and $K_3$ are positive constants independent of $n_2$ and $\gamma_i$. Here, the first term in \eqref{eq:step2U} follows from the Chernoff bound, and the second term in \eqref{eq:step2U} follows by writing
    \begin{align}
        &\Prob{\imath_{\mr{U}}(\mb{c}^n(2); Y^n) > \gamma_i} \notag \\
        &= \Prob{\hat{\rho}_{\mb{c}^n(2) Y^n} > a_i} \\
        &\leq \frac{K_4}{\sqrt{n}} \exp\left\{\frac{n}{2} \log (1 - a_i^2)\right\} (1 + o(1)), \label{eq:rhostep}
    \end{align}
    where $a_i \in (0, 1)$ satisfies $\gamma_i = -\frac{n}{2} \log(1-a_i^2)$, and $K_4$ is a positive constant that depends only on $a_i$. The bound in \eqref{eq:rhostep} follows from \lemref{lem:rho}.

    \item Let 
    \begin{align}
        \imath^*(b_{xy}, b_x, b_y) \triangleq -\frac{1}{2} \log \left( 1 - \frac{b_{xy}^2}{b_x b_y} \right).
    \end{align}
    Then, we have
    \begin{align}
        \imath_{\mr{U}}(X^n; Y^n) = n \, \imath^*\left(\frac{1}{n} \sum_{i = 1}^n X_i Y_i, \frac{1}{n} \sum_{i = 1}^n X_i^2, \frac{1}{n} \sum_{i = 1}^n Y_i^2 \right). 
    \end{align}
    Hence, $\imath_{\mr{U}}(X^n; Y^n)$ satisfies the conditions to apply \lemref{lem:nonlinear}. Taking the second-order Taylor series expansion of $\imath^*(\cdot)$ around $(P, P, P + \sigma_0^2)$, we get
    \begin{align}
        &\imath_{\mr{U}}(X^n; Y^n) \notag \\
        &= \imath(X^n; Y^n) + \frac{1}{2} W_n^\top \nabla^2 \imath^*(P, P, P + \sigma_0^2) W_n + o(1),
    \end{align}
    where
    \begin{align}
        W_n = \frac{1}{\sqrt{n}} \sum_{i = 1}^n \left(X_i Y_i - P, X_i^2 - P, Y_i^2 - (P + \sigma_0^2) \right),
    \end{align}
    and
    \begin{align}
        \imath(X^n; Y^n) = n C(S) - \sum_{i = 1}^n \frac{(Y_i - X_i)^2}{2 \sigma_0^2} + \sum_{i = 1}^n \frac{Y_i^2}{2 (P + \sigma_0^2)}
    \end{align}
    is the information density associated with the Gaussian channel under the Gaussian input distribution $P_X = \mc{N}(0, P)$. This means that for $(X^n, Y^n) \sim P_X^n P_{Y|X}^n$, similar to the DMC case, the information density $\imath(X^n; Y^n)$ is the first-order Taylor series approximation to the universal metric $\imath_{\mr{U}}(X^n; Y^n)$. 
    Therefore, applying \lemref{lem:nonlinear}, we get that 
    \begin{align}
        \E{\tau^{(1)}} &\leq \frac{\gamma_1}{C(S)} + O(1) \\
        \E{\tau^{(2)} - \tau^{(1)}} &\leq \frac{\gamma_2 - \gamma_1}{C(S)} + O(1).
    \end{align}

    \item We set the parameters of the HT phase as
    \begin{align}
        a_{\mr{A}} = a_{\mr{R}} = \log n_1.
    \end{align}
    Using \eqref{eq:Pcond}, we check that 
    \begin{align}
        D(P_{Y|X = x_{\mr{A}}} \| P_{Y|X = x_{\mr{R}}}) &= D(\tilde{P}_{Y|X = x_{\mr{A}}} \| \tilde{P}_{Y|X = x_{\mr{R}}}) \notag \\
        &\quad + O\left( \sqrt{\frac{\log n_1}{n_1}} \right). \label{eq:Dgap}
    \end{align}
    \eqref{eq:newHT}--\eqref{eq:newERA2} hold. 
    Combining \eqref{eq:THTA}--\eqref{eq:THTR} with \eqref{eq:Dgap}, we get
    \begin{align}
         \E{\tau_{\mr{HT}} | H_{\mr{A}}, \mc{G}} &=  \E{\tau_{\mr{HT}} | H_{\mr{R}}, \mc{G}} \\
         &\leq \frac{a_{\mr{A}} \left(1 + O\left(\sqrt{\frac{\log n_1}{n_1}}\right)\right)}{D(P_{Y|X = x_{\mr{A}}} \| P_{Y|X = x_{\mr{R}}})} + O(1) \label{eq:newHT} \\
        &= \frac{a_{\mr{A}} }{D(P_{Y|X = x_{\mr{A}}} \| P_{Y|X = x_{\mr{R}}})} + O(1). 
    \end{align}
    Combining \eqref{eq:EAR}--\eqref{eq:ERA} with \eqref{eq:Dgap}, we get
    \begin{align}
        \Prob{\mc{E}_{\mr{A} \to \mr{R}} | \mc{G}} &= \Prob{\mc{E}_{\mr{R} \to \mr{A}} | \mc{G}} \\
        &\leq \exp\left\{- a_{\mr{A}} \left(1 + O\left(\sqrt{\frac{\log n_1}{n_1}}\right)\right)\right\} \\
        &= \frac{1}{n_1} (1 + o(1)) \label{eq:newERA2} \\
        \E{\tau_{\mr{HT}}} &\leq \Prob{\mc{G}^{\mr{c}}} \E{\tau_{\mr{HT}} | \mc{G}^{\mr{c}}}  + \E{\tau_{\mr{HT}} |  \mc{G}} \\
        &\leq \frac{1}{n_1} + \frac{\log n_1}{C_1(S)} + O(1) \\
        &= \frac{\log n_1}{C_1(S)} + O(1).
    \end{align}

    \item Lastly, the error probability of our UVLF code given that the code is not stopped at time zero is bounded as
    \begin{align}
       \epsilon' \leq \Prob{\mc{G}^{\mr{c}}} + \Prob{\mc{E}^{(1)}} \Prob{\mc{E}_{\mr{R} \to \mr{A}} | \mc{G}} + \Prob{\mc{E}^{(2)}}. \label{eq:unionUVLF}
    \end{align}

    Due to \eqref{eq:step2U}, we set the parameters in \eqref{eq:gammaM}--\eqref{eq:gamma2M} as
    \begin{align}
    \gamma_1 &= \log M + \frac{1}{2} \log n_1 + (1 + \delta)\log \log n_1 \label{eq:gamma1new}\\
    \gamma_2 &= \log M + \frac{3}{2} \log n_1 + \delta \log \log n_1. \label{eq:gamma2new}
\end{align} 
With the choice of parameters in \eqref{eq:gamma1new} and \eqref{eq:gamma2new}, we have
\begin{align}
    \epsilon' \leq \frac{1}{n_1}
\end{align}
for large enough $M$. We choose the parameter $\epsilon_0$ as in \eqref{eq:neweps0} to ensure that the error probability does not exceed $\epsilon$.

We apply the bound in \eqref{eq:taubound} to get
\begin{align}
N' \leq \frac{\gamma_1}{C(S)} + \frac{\log n_1}{C_1(S)} + O(1).
\end{align}

Following the steps in \eqref{eq:logM1}--\eqref{eq:lastlogM}, we complete the proof of \eqref{eq:GaussianUVLF}.
\end{enumerate}

\section{Conclusion} \label{sec:conclusion}
In this work, we study variable-length feedback codes over known and unknown channels in the asymptotic regime that the error probability $\epsilon$ is non-vanishing as the average decoding time $N$ approaches infinity. 

Our achievability bound for both VLF codes employs a modified Yamamoto--Itoh scheme that has two communication phases and one confirmation phase, where each phase has a random length that depends on the noise realization. We also employ the stop-at-time-zero strategy used in \cite{polyanskiy2011feedback}, which enables to achieve the $\epsilon$-capacity of VLF codes. \thmref{thm:nonasymp} presents our novel non-asymptotic achievability bound for VLF codes. \thmref{thm:VLF} is our second-order achievability bound for VLF codes, which refines the second-order term achieved in \cite[Th.~2]{polyanskiy2011feedback} from $-\log N$ and $-\frac{C}{C_1} \log N$, where $C$ is the capacity, and $C_1$ is the optimal reliability function at zero rate. 

For UVLF codes, we develop a single-phase scheme that universalizes Polyanskiy \emph{et al.}'s scheme in \cite{polyanskiy2011feedback}. Similar to \cite{tchamkerten2006variable, goppa, lomnitzfeder}, we employ the empirical mutual information between the input and output sequences as our decoding metric. \thmref{thm:UVLF} presents our second-order achievability bound for UVLF codes over DMCs. In the proof of \thmref{thm:UVLF}, we use the asymptotic expansion in \cite[Th.~4.5]{woodroofebook} for the stopping time associated with a smooth function of an average of random vectors. In \lemref{lem:empMI}, we prove a tail probability bound with a refined pre-factor for the empirical mutual information evaluated on a joint type formed from two independent sequences, which plays a critical role in the derivation of the second-order term  in \thmref{thm:UVLF}. 

Our results extend to the Gaussian channel with known and unknown noise variances and an average power constraint. \thmref{thm:VLF_Gaussian} is our achievability bound for VLF codes over the Gaussian channel, which refines the bound in \cite[Th.~1]{truong2016gaussian}. For UVLF codes over the Gaussian channel, similar to \cite{lomnitzfeder}, we employ the universal metric $-\frac{1}{2} \log (1 - \hat{\rho}^2_{X^n Y^n})$, where $\hat{\rho}_{X^n Y^n}$ is the empirical correlation coefficient between $X^n$ and $Y^n$; this metric corresponds to the mutual information of two jointly Gaussian random variables with the correlation coefficient $\hat{\rho}_{X^n Y^n}$. The fact that the unknown noise variance $\sigma_0^2$ can be reliably estimated solely by the received power $\frac{1}{n} \norm{Y^n}_2^2$ makes it possible to universalize the Yamamoto--Itoh code for the Gaussian channel with an unknown noise variance.

\appendices
\section{Proof of \lemref{lem:empMI}} \label{app:proof_EMI}
We bound $\Prob{n I(\hat{P}_{\bar{X}^n}, \hat{P}_{Y^n|\bar{X}^n}) \geq \gamma}$ from above by two different approaches. We have 
    \begin{align}
        &\Prob{n I(\hat{P}_{\bar{X}^n}, \hat{P}_{Y^n|\bar{X}^n}) \geq \gamma} \notag \\
        &\leq \mathbb{P} \bigg[D(\hat{P}_{\bar{X}^n Y^n} \| P_X P_Y) \notag \\
        &\quad \geq \inf\limits_{Q_{XY} \colon n I(Q_X, Q_{Y|X}) \geq \gamma} D(Q_{XY} \| P_X P_Y) \bigg] \\
        &\leq \mathbb{P} \bigg[D(\hat{P}_{\bar{X}^n Y^n} \| P_X P_Y) \notag \\
        &\quad \geq \inf\limits_{Q_{XY} \colon n I(Q_X, Q_{Y|X}) \geq \gamma}  I(Q_X, Q_{Y|X}) \bigg]\label{eq:step1mardia} \\
        &= \Prob{D(\hat{P}_{\bar{X}^n Y^n} \| P_X P_Y) \geq \frac{\gamma}{n}} \label{eq:mot} \\
        &\leq c_1 \sum_{i = 1}^{|\mc{X}||\mc{Y}|-2} \left(\left(c_0 \frac{n}{i}\right)^{\frac{i}{2}} + 1 \right)  \exp\{-\gamma\}, \label{eq:mardia2020}
    \end{align}
    where $c_0 \approx 3.1967$ and $c_1 \approx 2.9290$. Inequality \eqref{eq:step1mardia} follows from $D(Q_{XY} \| P_X P_Y) = I(Q_X, Q_{Y|X}) + D(Q_X \| P_X) + D(Q_Y \| P_Y)$ and the non-negativity of the KL divergence. Inequality \eqref{eq:mardia2020} follows from the novel method of types bound from \cite[Th.~3]{mardia2020}. Since the prefactor in \eqref{eq:mardia2020} is $O(n^{\frac{|\mc{X}||\mc{Y}|-2}{2}})$, \eqref{eq:I1} follows, where $d$ is replaced with the first argument in the minimum in \eqref{eq:I2}. Note that the standard method of types from \cite[Lemma II.1]{csiszarMethod} bounds \eqref{eq:mot} by $(n+1)^{|\mc{X}||\mc{Y}|-1} \exp\{-\gamma\}$.

    To show \eqref{eq:I1} with $d$ replaced with the second argument in the minimum in \eqref{eq:I2}, we apply the Chernoff bound and get
    \begin{align}
        &\Prob{n I(\hat{P}_{\bar{X}^n}, \hat{P}_{Y^n|\bar{X}^n}) \geq \gamma} \notag \\
        &\quad \leq \E{\exp\{n I(\hat{P}_{\bar{X}^n}, \hat{P}_{ Y^n|\bar{X}^n}) \}} \exp\{-\gamma\}. \label{eq:chernoff}
    \end{align}
    
    Noting that $\log P_X^n(x^n) = \sum_{x \in \mc{X}} {\hat{P}_{x^n}(x)} \log P_X(x)$, we write the expectation in \eqref{eq:chernoff} as
    \begin{align}
        &\E{\exp\{n I(\hat{P}_{\bar{X}^n}, \hat{P}_{Y^n|\bar{X}^n}) \}} \notag \\
        &= \sum_{x^n, y^n} P_X^n(x^n) P_Y^n(y^n) \exp\{n I(\hat{P}_{x^n}, \hat{P}_{y^n|x^n})\} \\
        &= \sum_{x^n, y^n} \exp\left\{-n \left(D(\hat{P}_{x^n} \| P_X) + D(\hat{P}_{y^n} \| P_Y) + H(\hat{P}_{x^n y^n}) \right) \right\} \\
        & \sum_{Q_{XY} \in \mc{P}_n(\mc{X}, \mc{Y})} |\mc{T}_n(Q_{XY})| \notag \\
        &\quad \exp\left\{-n \left(D(Q_X \| P_X) + D(Q_Y \| P_Y) + H(Q_{XY}) \right) \right\}. \label{eq:expQ}
    \end{align}
    
    Next, we use the tight bound on the size of the type class \cite[Exercise 2.2]{csiszarbook}
    \begin{align}
        |\mc{T}_n(Q_{X})| \leq \exp\{n H(Q_X)\} (2 \pi n)^{-\frac{|\mc{X}|-1}{2}} \prod_{x \in \mc{X}} \frac{1}{\sqrt{\tilde{Q}_X(x)}}, \label{eq:typeclass}
    \end{align}
    where $\tilde{Q}_X(x) = \frac{1}{2 \pi n}$ if $Q_X(x) = 0$ and  $\tilde{Q}_X(x) = Q_X(x)$ otherwise.

    Applying \eqref{eq:typeclass} to \eqref{eq:expQ}, we get
    \begin{align}
        &\E{\exp\{n I(\hat{P}_{\bar{X}^n}, \hat{P}_{Y^n|\bar{X}^n}) \}} =  (2 \pi n)^{-\frac{|\mc{X}||\mc{Y}|-1}{2}}  \notag \\
        &  \quad \sum_{Q_{XY} \in \mc{P}_n(\mc{X}, \mc{Y})} \Bigg[ \exp\left\{-n \left(D(Q_X \| P_X) + D(Q_Y \| P_Y) \right) \right\} \notag \\
        &\quad \prod \limits_{(x, y) \in \mc{X} \times \mc{Y}} \frac{1}{\sqrt{\tilde{Q}_{XY}(x, y)}} \Bigg] . \label{eq:exp2}
    \end{align}
    Define the sets
    \begin{align}
        \mc{A}_n(Q_X, Q_Y) \triangleq \{ V_{XY} \in \mc{P}_n(\mc{X}, \mc{Y}) \colon V_X = Q_X, V_Y = Q_Y\}. 
    \end{align}
    
    We rewrite the summation in \eqref{eq:exp2} to get
    \begin{align}
        &\E{\exp\{n I(\hat{P}_{\bar{X}^n}, \hat{P}_{Y^n|\bar{X}^n}) \}} \leq  (2 \pi n)^{-\frac{|\mc{X}||\mc{Y}|-1}{2}}  \notag \\
        &  \quad \left(\sum_{\substack{Q_{X} \in \mc{P}_n(\mc{X}) \\ Q_{Y} \in \mc{P}_n(\mc{Y}) }}  \exp\left\{-n \left(D(Q_X \| P_X) + D(Q_Y \| P_Y) \right) \right\} \right) \label{eq:laplace}  \\
        &\quad \left(\max \limits_{\substack{V_X \in \mc{P}(\mc{X}) \\ V_Y \in \mc{P}(\mc{Y})} } \sum_{V_{XY} \in \mc{A}_n(V_X, V_Y)} \prod \limits_{(x, y) \in \mc{X} \times \mc{Y}} \frac{1}{\sqrt{\tilde{V}_{XY}(x, y)}} \right).  \label{eq:sumint}
    \end{align}
    Note that $|\mc{A}_n(Q_X, Q_Y)| \leq (n+1)^{(|\mc{X}|-1)(|\mc{Y}|-1)}$. Bounding the summation in \eqref{eq:sumint} by an appropriate integral, we get
    \begin{align}
       &\max \limits_{\substack{V_X \in \mc{P}(\mc{X}) \\ V_Y \in \mc{P}(\mc{Y})} } \sum_{V_{XY} \in \mc{A}_n(V_X, V_Y)} \prod \limits_{(x, y) \in \mc{X} \times \mc{Y}} \frac{1}{\sqrt{\tilde{V}_{XY}(x, y)}} \notag \\
       &\leq C_2 (n+1)^{(|\mc{X}|-1)(|\mc{Y}|-1)}, \label{eq:C2}
    \end{align}
    where $C_2 > 0$ is a constant depending on $|\mc{X}|$ and $|\mc{Y}|$. It only remains to bound the summation in \eqref{eq:laplace}. To do that, we use the following asymptotic result, which can be viewed as Laplace's method for sums over types.
    \begin{lemma} \label{lem:laplace}
        Let $f \colon \mc{P}(\mc{X}) \to \mathbb{R}$ be a function with a unique minimum at $P_X^*$. Let $\epsilon > 0$ and let $B_{\epsilon}$ be a ball of radius $\epsilon$ centered at $P_X^*$. Assume that the derivatives of $f$ up to third order exist and are bounded in $B_{\epsilon}$. Assume that the minimum eigenvalue of  $\nabla^2f(P_X)$ is bounded below by 0 for all $P_X \in B_{\epsilon}$. Then,
        \begin{align}
            &\sum_{P_X \in \mc{P}_n(\mc{X})} \exp\{- n f(P_X)\} \notag \\
            &= (2 \pi n)^{\frac{|\mc{X}|-1}{2}} \exp\{- n f(P_X^*)\} \frac{1}{\sqrt{\det(\nabla^2f(P_X^*))}} (1 + o(1)).
        \end{align}
    \end{lemma}

    The function $f(\cdot) = D(\cdot \| P_X)$ satisfies the conditions of \lemref{lem:laplace} given that $P_X(x) > 0$ for all $x \in \mc{X}$ with the minimizer $P_X^* = P_X$ and the minimum value of zero. Therefore, applying  \lemref{lem:laplace} to \eqref{eq:exp2} twice, we get
    \begin{align}
        &\sum_{\substack{Q_{X} \in \mc{P}_n(\mc{X}) \\ Q_{Y} \in \mc{P}_n(\mc{Y}) }}  \exp\left\{-n \left(D(Q_X \| P_X) + D(Q_Y \| P_Y) \right) \right\} \notag \\
        &\leq C_3 (n + 1)^{\frac{(|\mc{X}|-1)(|\mc{Y}|-1)|}{2}} (1 + o(1)), \label{eq:C3}
    \end{align}
    where $C_3 > 0$ is a constant. Finally, combining \eqref{eq:chernoff}, \eqref{eq:laplace}--\eqref{eq:C2}, and \eqref{eq:C3} completes the proof.

    \section{Proof of \eqref{eq:etastep}} \label{app:eta}
    Define 
    \begin{align}
        Q_i \triangleq  \mr{Bernoulli}(i / n), \quad i = 0, \dots, n.
    \end{align}
    By changing measure from $P_X P_{Y|X}$ to $P_X P_Y$, we get
    \begin{align}
        \E{\exp\{\eta_n\}} &= \sum_{z^n \in \{0, 1\}^n} \left(\frac{1}{2} \right)^n \exp\{ n (\log 2 - H(\hat{P}_{z^n}))\} \\
        &= \sum_{z^n \in \{0, 1\}^n} \exp\{- n H(\hat{P}_{z^n}))\} \\
        &= \sum_{i = 0}^n |\mc{T}_n(Q_i)| \exp\{- n H(Q_i)\} \label{eq:group}\\
        &= \sum_{i = 0}^n \frac{1}{\sqrt{2 \pi n}} \prod_{x \in \{0, 1\}} \frac{1}{\sqrt{\tilde{Q}_i(x)}} \label{eq:Tstep}\\
        &\leq K_2 n,
    \end{align}
    where $K_2$ is a positive constant, \eqref{eq:group} follows by grouping the sequences of the same type, and \eqref{eq:Tstep} follows from \eqref{eq:typeclass}.
        
\bibliographystyle{IEEEtran}
\bibliography{mac} 

\begin{IEEEbiographynophoto}{Recep Can Yavas}
(Member, IEEE) received the B.S. degree (Hons.) in electrical engineering from Bilkent University, Ankara, Turkey, in 2016. He received the M.S. and Ph.D. degrees in electrical engineering from California Institute of Technology (Caltech) in 2017 and 2023, respectively. He was a research fellow at CNRS@CREATE, Singapore, between 2022 and 2024. In October 2024, he joined the Department of Computer Science at National University of Singapore as a research fellow. 
His research interests include information theory, communications, and multi-armed bandits.
\end{IEEEbiographynophoto}

\begin{IEEEbiographynophoto}{Vincent Y. F. Tan} (Senior Member, IEEE) was born in Singapore, in 1981.
He received the B.A. and M.Eng. degrees in electrical and information
science from Cambridge University in 2005 and the Ph.D. degree in electrical
engineering and computer science (EECS) from Massachusetts Institute of
Technology (MIT) in 2011. He is currently a Professor with the Department of Mathematics and the
Department of Electrical and Computer Engineering (ECE), National University of Singapore (NUS). His research interests include information theory,
machine learning, and statistical signal processing. 

He is an elected member
of the IEEE Information Theory Society Board of Governors. He was an
IEEE Information Theory Society Distinguished Lecturer from 2018 to 2019.
He received the MIT EECS Jin-Au Kong Outstanding Doctoral Thesis Prize
in 2011, the NUS Young Investigator Award in 2014, Singapore National
Research Foundation (NRF) Fellowship (Class of 2018), and the NUS Young
Researcher Award in 2019. He is also serving as a Senior Area Editor for
{\em IEEE Transactions on Signal Processing} and an Associate Editor in
Machine Learning and Statistics for {\em  IEEE Transactions on  Information Theory}. He also regularly serves as the Area Chair for Prominent
Machine Learning Conferences, such as the International Conference on
Learning Representations (ICLR) and the Conference on Neural Information
Processing Systems (NeurIPS).
\end{IEEEbiographynophoto}

\end{document}